\documentclass[10pt,journal,compsoc]{IEEEtran}

\usepackage{amsmath}
\usepackage{amssymb}
\usepackage{amsfonts}
\usepackage{graphicx}
\usepackage{epsfig}
\usepackage{cite}
\usepackage{multirow}
\usepackage{subfigure}
\usepackage{xcolor}
\usepackage[ruled,vlined,lined,ruled,linesnumbered]{algorithm2e}
\usepackage{url}

\begin{document}

\title{Deep Reinforcement Learning for Online Computation Offloading in Wireless Powered Mobile-Edge Computing Networks}

\author{Liang~Huang,~\IEEEmembership{Member,~IEEE,}
        Suzhi~Bi,~\IEEEmembership{Senior Member,~IEEE,}
        and~Ying-Jun~Angela~Zhang,~\IEEEmembership{Senior Member,~IEEE}% <-this % stops a space
\IEEEcompsocitemizethanks{\IEEEcompsocthanksitem L.~Huang is with the College of Information Engineering, Zhejiang University of Technology, Hangzhou, China 310058, (e-mail: lianghuang@zjut.edu.cn).\protect\\
\IEEEcompsocthanksitem S.~Bi is with the College of Electronic and Information Engineering, Shenzhen University, Shenzhen, Guangdong, China 518060 (e-mail: bsz@szu.edu.cn).\protect\\
\IEEEcompsocthanksitem Y-J.~A.~Zhang is with the Department of Information Engineering, The Chinese University of Hong Kong, Shatin, N.T., Hong Kong. (e-mail: yjzhang@ie.cuhk.edu.hk).}}

\IEEEtitleabstractindextext{
\begin{abstract}
Wireless powered mobile-edge computing (MEC) has recently emerged as a promising paradigm to enhance the data processing capability of low-power networks, such as wireless sensor networks and internet of things (IoT). In this paper, we consider a wireless powered MEC network that adopts a binary offloading policy, so that each computation task of wireless devices (WDs) is either executed locally or fully offloaded to an MEC server. Our goal is to acquire an online algorithm that optimally adapts task offloading decisions and wireless resource allocations to the time-varying wireless channel conditions. This requires quickly solving hard combinatorial optimization problems within the channel coherence time, which is hardly achievable with conventional numerical optimization methods. To tackle this problem, we propose a Deep Reinforcement learning-based Online Offloading (DROO) framework that implements a deep neural network as a scalable solution that learns the binary offloading decisions from the experience. It eliminates the need of solving combinatorial optimization problems, and thus greatly reduces the computational complexity especially in large-size networks. To further reduce the complexity, we propose an adaptive procedure that automatically adjusts the parameters of the DROO algorithm on the fly. Numerical results show that the proposed algorithm can achieve near-optimal performance while significantly decreasing the computation time by more than an order of magnitude compared with existing optimization methods. For example, the CPU execution latency of DROO is less than $0.1$ second in a $30$-user network, making real-time and optimal offloading truly viable even in a fast fading environment.
\end{abstract}

}

\maketitle

\IEEEdisplaynontitleabstractindextext

\IEEEpeerreviewmaketitle

\IEEEraisesectionheading{\section{Introduction}\label{sec:introduction}}
\IEEEPARstart{D}{ue} to the small form factor and stringent production cost constraint, modern Internet of Things (IoT) devices are often limited in battery lifetime and computing power. Thanks to the recent advance in \emph{wireless power transfer} (WPT) technology, the batteries of {wireless devices (WDs)} can be continuously charged over the air without the need of battery replacement\cite{bi2015mcom}. Meanwhile, the device computing power can be effectively enhanced by the recent development of \emph{mobile-edge computing} (MEC) technology\cite{chiang2016iotj, mao2016jsac}. With MEC, the WDs can offload computationally intensive tasks to nearby edge servers to reduce computation latency and energy consumption\cite{chen2016ton, you2017wcom}.

\begin{figure}
    \centering
    \begin{center}
        \includegraphics[width=0.47\textwidth]{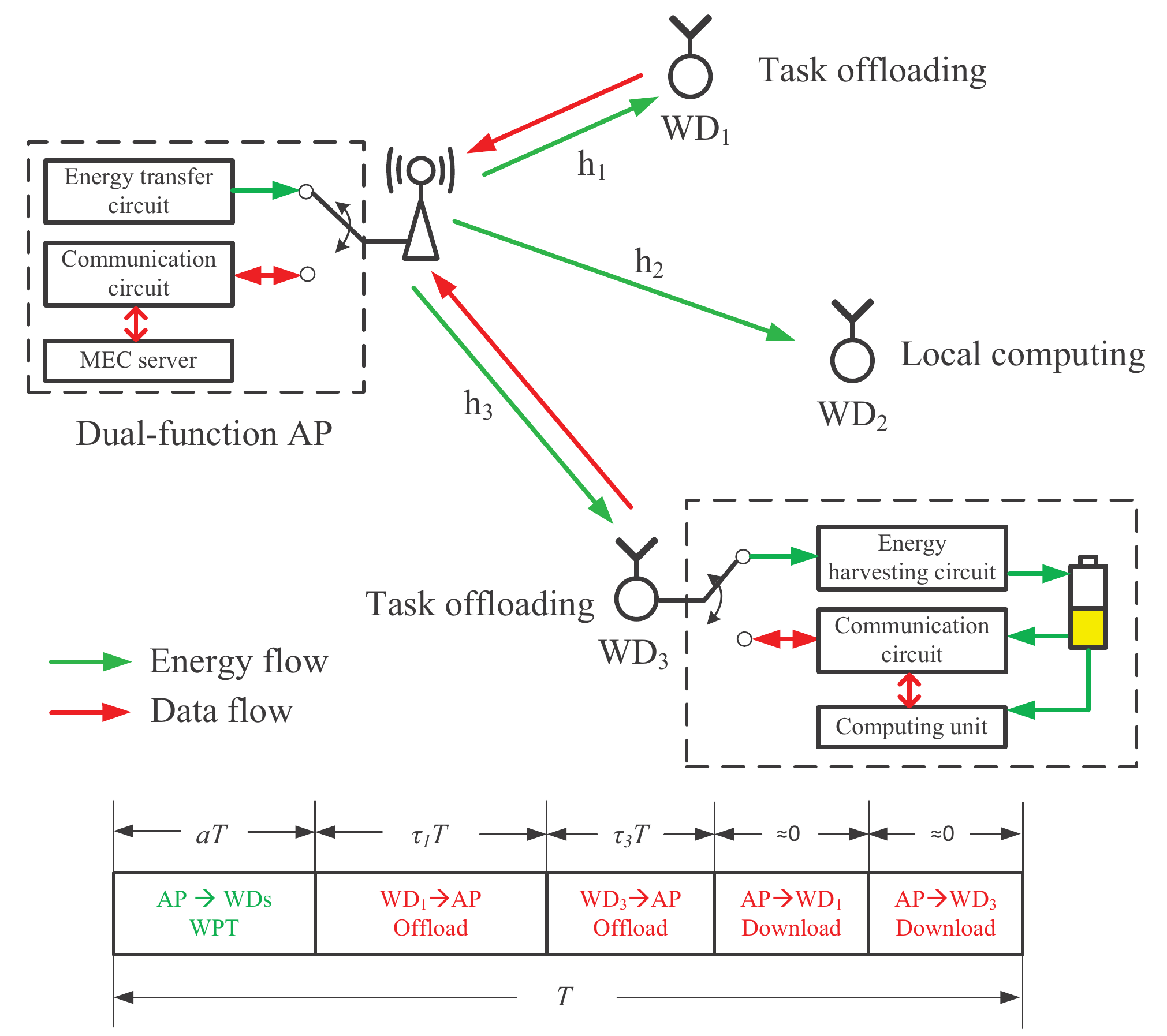}
    \end{center}
    \caption{An example of the considered wireless powered MEC network and system time allocation.}
    \label{fig:model}
\end{figure}

The newly emerged \emph{wireless powered MEC} combines the advantages of the two aforementioned technologies, and thus holds significant promise to solve the two fundamental performance limitations for IoT devices \cite{wang2018joint, bi2018twc}. In this paper, we consider a wireless powered MEC system as shown in Fig.~\ref{fig:model}, where the access point (AP) is responsible for both transferring RF (radio frequency) energy to and receiving computation offloading from the WDs. In particular, the WDs follow a \textit{binary task offloading} policy\cite{mao2017survey}, where a task is either computed locally or offloaded to the MEC server for remote computing. The system setup may correspond to a typical outdoor IoT network, where each energy-harvesting wireless sensor computes a non-partitionable simple sensing task with the assistance of an MEC server.

In a wireless fading environment, the time-varying wireless channel condition largely impacts the optimal offloading decision of a wireless powered MEC system \cite{2016:You}. In a multi-user scenario, a major challenge is the joint optimization of individual computing mode (i.e., offloading or local computing) and wireless resource allocation (e.g., the transmission air time divided between WPT and offloading). Such problems are generally formulated as mixed integer programming (MIP) problems due to the existence of binary offloading variables. To tackle the MIP problems, branch-and-bound algorithms \cite{narendra1977tc} and dynamic programming \cite{bertsekas1995dynamic} have been adopted, however, with prohibitively high computational complexity, especially for large-scale MEC networks. To reduce the computational complexity, heuristic local search \cite{bi2018twc,tran2017arxiv}  and convex relaxation \cite{yang2016infocom,tony2017toc} methods are proposed. However, both of them require considerable number of iterations to reach a satisfying local optimum. Hence, they are not suitable for making real-time offloading decisions in fast fading channels, as the optimization problem needs to be re-solved once the channel fading has varied significantly.

In this paper, we consider a wireless powered MEC network with one AP and multiple WDs as shown in Fig.~\ref{fig:model}, where each WD follows a binary offloading policy. In particular, we aim to jointly optimize the individual WD's task offloading decisions, transmission time allocation between WPT and task offloading, and time allocation among multiple WDs according to the time-varying wireless channels. Towards this end, we propose a deep reinforcement learning-based online offloading (DROO) framework to maximize the weighted sum of the computation rates of all the WDs, i.e., the number of processed bits within a unit time. Compared with the existing integer programming and learning-based methods, we have the following novel contributions:
\begin{enumerate}
  \item The proposed DROO framework learns from the past offloading experiences under various wireless fading conditions, and automatically improves its action generating policy. As such, it completely removes the need of solving complex MIP problems, and thus, the computational complexity does not explode with the network size.
  \item {Unlike many existing deep learning methods that optimize all system parameters at the same time resulting infeasible solutions, DROO decomposes the original optimization problem into an offloading decision sub-problem and a resource allocation sub-problem, such that all physical constraints are guaranteed. It works for continuous state spaces and does not require the discretization of channel gains, thus, avoiding the curse of dimensionality problem.}
  \item To efficiently generate offloading actions, we devise a novel order-preserving action generation method. Specifically, it only needs to select from few candidate actions each time, thus is computationally feasible and efficient in large-size networks with high-dimensional action space. Meanwhile, it also provides high diversity in the generated actions and leads to better convergence performance than conventional action generation techniques.
  \item We further develop an adaptive procedure that automatically adjusts the parameters of the DROO algorithm on the fly. Specifically, it gradually decreases the number of convex resource allocation sub-problems to be solved in a time frame. This effectively reduces the computational complexity without compromising the solution quality.
\end{enumerate}
We evaluate the proposed DROO framework under extensive numerical studies. Our results show that on average the DROO algorithm achieves over $99.5\%$ of the computation rate of the existing near-optimal benchmark method \cite{bi2018twc}. Compared to the Linear Relaxation (LR) algorithm \cite{yang2016infocom}, it significantly reduces the CPU execution latency by more than an order of magnitude, e.g., from $0.81$ second to $0.059$ second in a $30$-user network. This makes real-time and optimal design truly viable in wireless powered MEC networks even in a fast fading environment. {The complete source code implementing DROO is available at \url{https://github.com/revenol/DROO}.}

The remainder of this paper is organized as follows. In Section~\ref{sec:relatedwork}, a review of related works in literature is presented. In Section~\ref{sec:model}, we describe the system model and problem formulation. We introduce the detailed designs of the DROO algorithm in Section~\ref{sec:DRL}. Numerical results are presented in Section~\ref{sec:result}. Finally, the paper is concluded in Section \ref{sec:conclusion}.

\section{{Related Work}} \label{sec:relatedwork}
There are many related works that jointly model the computing mode decision problem and resource allocation problem in MEC networks as the MIP problems. For instance, \cite{bi2018twc} proposed a coordinate descent (CD) method that searches along one variable dimension at a time. \cite{tran2017arxiv} studies a similar heuristic search method for multi-server MEC networks, which iteratively adjusts binary offloading decisions. Another widely adopted heuristic is through convex relaxation, e.g., by relaxing integer variables to be continuous between $0$ and $1$ \cite{yang2016infocom} or by approximating the binary constraints with quadratic constraints \cite{tony2017toc}. Nonetheless, on one hand, the solution quality of the reduced-complexity heuristics is not guaranteed. On the other hand, both search-based and convex relaxation methods require considerable number of iterations to reach a satisfying local optimum and are inapplicable for fast fading channels.

Our work is inspired by recent advantages of deep reinforcement learning in handling reinforcement learning problems with large state spaces \cite{mnih2015nature} and action spaces \cite{dulac2015action}. In particular, it relies on deep neural networks (DNNs) \cite{lecun2015nature} to learn from the training data samples, and eventually produces the optimal mapping from the state space to the action space. There exists limited work on deep reinforcement learning-based offloading for MEC networks \cite{zhao2017dqn,min2017learningarxiv,zhang2018DQNarxiv,huang2018monet,huang2019dcn}. By taking advantage of parallel computing, \cite{huang2018monet} proposed a distributed deep learning-based offloading (DDLO) algorithm for MEC networks. For an energy-harvesting MEC networks, \cite{min2017learningarxiv} proposed a deep Q-network (DQN) based offloading policy to optimize the computational performance. Under the similar network setup, \cite{zhang2018DQNarxiv} studied an online computation offloading policy based on DQN under random task arrivals. However, both DQN-based works take discretized channel gains as the input state vector, and thus suffer from the curse of dimensionality and slow convergence when high channel quantization accuracy is required. Besides, because of its exhaustive search nature in selecting the action in each iteration, DQN is not suitable for handling problems with high-dimensional action spaces \cite{lillicrap2015continuous}. In our problem, there are a total of $2^N$ offloading decisions (actions) to choose from, where DQN is evidently inapplicable even for a small $N$, e.g., $N=20$.

\section{Preliminary}\label{sec:model}
\subsection{System Model}
As shown in Fig.~\ref{fig:model}, we consider a wireless powered MEC network consisting of an AP and $N$ fixed WDs, denoted as a set $\mathcal{N}=\{1,2,\dots,N\}$, where each device has a single antenna. In practice, this may correspond to a static sensor network or a low-power IoT system. The AP has stable power supply and can broadcast RF energy to the WDs. Each WD has a rechargeable battery that can store the harvested energy to power the operations of the device. Suppose that the AP has higher computational capability than the WDs, so that the WDs may offload their computing tasks to the AP. Specifically, we suppose that WPT and communication (computation offloading) are performed in the same frequency band. Accordingly, a time-division-multiplexing (TDD) circuit is implemented at each device to avoid mutual interference between WPT and communication.

The system time is divided into consecutive time frames of equal lengths $T$, {which is set smaller than the channel coherence time, e.g., in the scale of several seconds \cite{bultitude1987measurement,howard1990Doppler,herbert2014Characterizing} in a static IoT environment}. At each tagged time, both the amount of energy that a WD harvests from the AP and the communication speed between them are related to the wireless channel gain. Let $h_i$ denote the wireless channel gain between the AP and the $i$-th WD at a tagged time frame. The channel is assumed to be reciprocal in the downlink and uplink,\footnote{The channel reciprocity assumption is made to simplify the notations of channel state. However, the results of this paper can be easily extended to the case with unequal uplink and downlink channels.} and remain unchanged within each time frame, but may vary across different frames. At the beginning of a time frame, $aT$ amount of time is used for WPT, $a\in[0,1]$, where the AP broadcasts RF energy for the WDs to harvest. Specifically, the $i$-th WD harvests $E_i = \mu P h_i a T$ amount of energy, where $\mu\in(0,1)$ denotes the energy harvesting efficiency and $P$ denotes the AP transmit power \cite{bi2015mcom}. With the harvested energy, each WD needs to accomplish a prioritized computing task before the end of a time frame. {A unique weight $w_i$ is assigned to the $i$-th WD. The greater the weight $w_i$, the more computation rate is allocated to the $i$-th WD.} In this paper, we consider a binary offloading policy, such that the task is either computed locally at the WD (such as WD2 in Fig.~\ref{fig:model}) or offloaded to the AP (such as WD1 and WD3 in Fig.~\ref{fig:model}). Let $x_{i} \in \{0,1\}$ be an indicator variable, where $x_i=1$ denotes that the $i$-th user's computation task is offloaded to the AP, and $x_i=0$ denotes that the task is computed locally.

\subsection{Local Computing Mode}
A WD in the local computing mode can harvest energy and compute its task simultaneously \cite{wang2018joint}. Let $f_i$ denote the processor's computing speed (cycles per second) and $0\leq t_i\leq T$ denote the computation time. Then, the amount of processed bits by the WD is $f_i t_i/\phi$, where $\phi>0$ denotes the number of cycles needed to process one bit of task data. Meanwhile, the energy consumption of the WD due to the computing is constrained by $k_i f_i^3 t_i \leq E_i$, where $k_i$ denotes the computation energy efficiency coefficient \cite{yang2016infocom}. It can be shown that to process the maximum amount of data within $T$ under the energy constraint, a WD should exhaust the harvested energy and compute throughout the time frame, i.e., $t_i^*=T$ and accordingly $f_i^*=\left(\frac{E_i}{k_i T}\right)^{\frac{1}{3}}$. Thus, the local computation rate (in bits per second) is
\begin{equation}
\label{equ:r_li}
r_{L,i}^*(a)  = \frac{f_i^* t_i^*}{\phi T} =\eta_1 \left(\frac{h_i}{k_i}\right)^{\frac{1}{3}} a^{\frac{1}{3}},
\end{equation}
where $\eta_1 \triangleq  \left( \mu P \right)^{\frac{1}{3}}/\phi$ is a fixed parameter.

\subsection{Edge Computing Mode}
Due to the TDD constraint, a WD in the offloading mode can only offload its task to the AP after harvesting energy. We denote $\tau_i T$ as the offloading time of the $i$-th WD, $\tau_i\in[0,1]$. Here, we assume that the computing speed and the transmit power of the AP is much larger than the size- and energy-constrained WDs, e.g., by more than three orders of magnitude \cite{2016:You,wang2018joint}. Besides, the computation feedback to be downloaded to the WD is much shorter than the data offloaded to the edge server. Accordingly, as shown in Fig.~\ref{fig:model}, we safely neglect the time spent on task computation and downloading by the AP, such that each time frame is only occupied by WPT and task offloading, i.e.,
\begin{equation}
\label{11}
\sum_{i=1}^N \tau_i + a \leq 1.
\end{equation}

To maximize the computation rate, an offloading WD exhausts its harvested energy on task offloading, i.e., $P_i^* = \frac{E_i}{\tau_i T}$. Accordingly, the computation rate equals to its data offloading capacity, i.e.,
\begin{equation}
\label{equ:r_oi}
r_{O,i}^*(a,\tau_i) = \frac{B\tau_i}{v_u}\log_2\left(1+\frac{\mu P a h_i^2}{ \tau_i N_0}\right),
\end{equation}
where $ B $ denotes the communication bandwidth and $ N_0 $ denotes the receiver noise power.

\subsection{Problem Formulation}
Among all the system parameters in (\ref{equ:r_li}) and (\ref{equ:r_oi}), we assume that only the wireless channel gains $\mathbf{h} = \{h_i| i\in \mathcal{N}\}$ are time-varying in the considered period, while the others (e.g., $w_i$'s and $k_i$'s) are fixed parameters. Accordingly, the weighted sum computation rate of the wireless powered MEC network in a tagged time frame is denoted as
\begin{equation*}
Q\left(\mathbf{h},\mathbf{x},\boldsymbol{\tau},a\right) \triangleq \sum_{i=1}^N w_i \left( (1-x_i) r_{L,i}^*(a)  + x_i r_{O,i}^*(a,\tau_i) \right),
\end{equation*}
where $\mathbf{x} = \{x_i| i\in \mathcal{N}\}$ and $\boldsymbol{\tau} = \{\tau_i| i\in \mathcal{N}\}$.

For each time frame with channel realization $\mathbf{h}$, we are interested in maximizing the weighted sum computation rate:
 \begin{subequations}
   \begin{align}
  (P1):\ Q^*\left(\mathbf{h}\right)=\ & \underset{\mathbf{x}, \boldsymbol{\tau},a}{\text{maximize}}  && Q\left(\mathbf{h},\mathbf{x},\boldsymbol{\tau},a\right)  \\
    & \text{subject to}  && \mathsmaller \sum_{i=1}^N \tau_i + a \leq 1, \label{equ:p1b}\\
    & && a \geq 0,\ \tau_i\geq 0, \ \forall i\in \mathcal{N}, \label{equ:p1c}\\
    & && x_i \in \{0,1\}. \label{e:binary_x-i}
   \end{align}
   \label{equ:p1}
\end{subequations}
We can easily infer that $\tau_i =0$ if $x_i = 0$, i.e., when the $i$-th WD is in the local computing mode.

\begin{figure}
\centering
\includegraphics[width=0.45\textwidth]{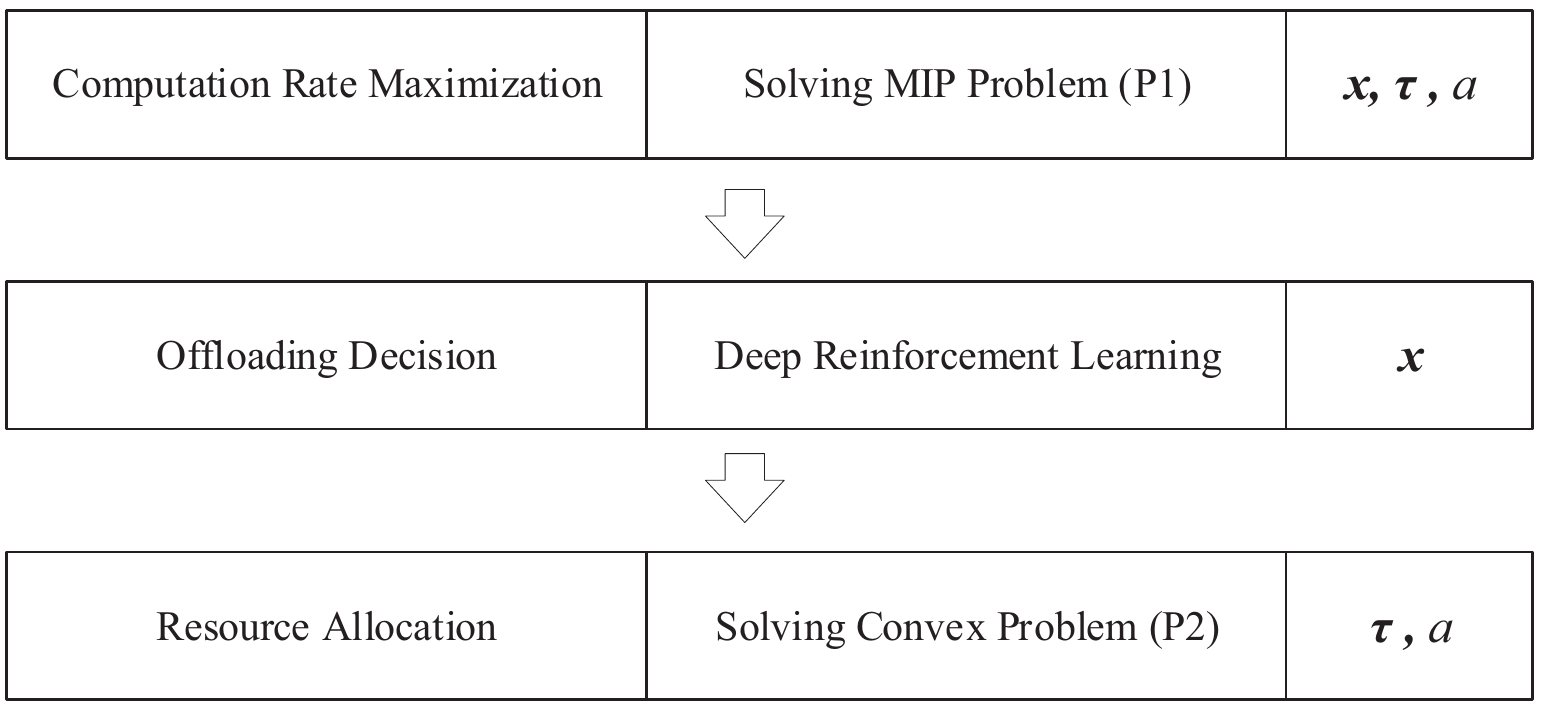}
\caption{The two-level optimization structure of solving (P1).}
\label{fig:problem}
\end{figure}

Problem (P1) is a mixed integer programming non-convex problem, which is hard to solve. However, once $\mathbf{x}$ is given, (P1) reduces to a convex problem as follows.
\begin{align*}
 (P2): \ Q^*\left(\mathbf{h},\mathbf{x}\right)=\ & \underset{ \boldsymbol{\tau}, a}{\text{maximize}} & &  Q\left(\mathbf{h},\mathbf{x},\boldsymbol{\tau},a\right) \\
    & \text{subject to} &  & \mathsmaller \sum_{i=1}^N \tau_i + a \leq 1,\\
    & & & a \geq 0,\ \tau_i\geq 0, \ \forall i\in \mathcal{N}.
\end{align*}
Accordingly, problem (P1) can be decomposed into two sub-problems, namely, offloading decision and resource allocation (P2), as shown in Fig.~\ref{fig:problem}:
\begin{itemize}
	\item \textit{Offloading Decision}: One needs to search among the $2^N$ possible offloading decisions to find an optimal or a satisfying sub-optimal offloading decision $\mathbf{x}$. For instance, meta-heuristic search algorithms are proposed in \cite{bi2018twc} and \cite{tran2017arxiv} to optimize the offloading decisions. However, due to the exponentially large search space, it takes a long time for the algorithms to converge.
  \item \textit{Resource Allocation}: The optimal time allocation $\left\{a^*,\boldsymbol{\tau}^*\right\}$ of the convex problem (P2) can be efficiently solved, e.g., using a one-dimensional bi-section search over the dual variable associated with the time allocation constraint in $O(N)$ complexity \cite{bi2018twc}.
\end{itemize}

The major difficulty of solving (P1) lies in the offloading decision problem. Traditional optimization algorithms require {iteratively adjusting the offloading decisions towards the optimum} \cite{bertsekas1995dynamic}, which is fundamentally infeasible for real-time system optimization under fast fading channel. To tackle the complexity issue, we propose a novel deep reinforcement learning-based online offloading (DROO) algorithm that can achieve a millisecond order of computational time in solving the offloading decision problem.

Before leaving this section, it is worth mentioning the advantages of applying deep reinforcement learning over supervised learning-based deep neural network (DNN) approaches (such as in \cite{hong2017spawc} and \cite{ye2018dlofdm}) in dynamic wireless applications. Other than the fact that deep reinforcement learning does not need manually labeled training samples (e.g., the $\left(\mathbf{h}, \mathbf{x}\right)$ pairs in this paper) as DNN, it is much more robust to the change of user channel distributions. For instance, the DNN needs to be completely retrained once some WDs change their locations significantly or are suddenly turned off. In contrast, the adopted deep reinforcement learning method can automatically update its offloading decision policy upon such channel distribution changes without manual involvement.  Those important notations used throughout this paper are summarized in Table~\ref{tab:notations}.
\begin{table}
    \caption{{Notations used throughout the paper}}
    \begin{center}
        \begin{tabular}{|p{0.4 in}| p{2.7in}|}
            Notation          &   Description     \\
            $N$          &  The number of WDs   \\
            $T$          &  The length of a time frame   \\
            $i$          &   Index of the $i$-th WD  \\
            $h_i$          &  The wireless channel gain between the $ i $-th WD and the AP   \\
            $a$          & The fraction of time that the AP broadcasts RF energy for the WDs to harvest    \\
            $E_i$          &  The amount of energy harvested by the $ i $-th WD  \\
            $P$          &  The AP transmit power when broadcasts RF energy    \\
            $\mu$          &  The energy harvesting efficiency   \\
            $w_i$          & The weight assigned to the $ i $-th WD    \\
            $x_i$          &  An offloading indicator for the $ i $-th WD   \\
            $f_i$          &  The processor's computing speed of the $ i $-th WD   \\
            $\phi$          &  The number of cycles needed to process one bit of task data   \\
            $t_i$          & The computation time of the $ i $-th WD    \\
            $k_i$          &  The computation energy efficiency coefficient   \\
            $\tau_i$          & The fraction of time allocated to the $ i $-th WD for task offloading     \\
            $B$          &  The communication bandwidth    \\
            $N_0$          &  The receiver noise power   \\
            $\mathbf{h}$          & The vector representation of  wireless channel gains $\{h_i| i\in \mathcal{N}\}$    \\
            $\mathbf{x} $          & The vector representation of offloading indicators  $\{x_i| i\in \mathcal{N}\} $   \\
            $\boldsymbol{\tau}$          &The vector representation of   $ \{\tau_i| i\in \mathcal{N}\}$   \\
            $Q(\cdot)$          & The weighted sum computation rate function    \\
            $\pi$          &  Offloading policy function   \\
            $\theta$          & The parameters of the DNN    \\
            $\hat{\mathbf{x}}_t$          &  Relaxed computation offloading action   \\
            $K$          & The number of quantized binary offloading actions    \\
            $g_K$          & The quantization function     \\
            $L(\cdot)$          & The training loss function of the DNN    \\
            $\delta$          &  The training interval of the DNN  \\
            $\Delta$          &  The updating interval for $ K $   \\
        \end{tabular}
    \end{center}
    \label{tab:notations}
\end{table}

\section{The DROO Algorithm}\label{sec:DRL}
We aim to devise an offloading policy function $\pi$ that quickly generates an optimal offloading action $\mathbf{x}^* \in \{0,1\}^N$ of (P1) once the channel realization $\mathbf{h} \in \mathbb{R}_{>0}^N$ is revealed at the beginning of each time frame. The policy is denoted as
\begin{align}
 \pi : \mathbf{h} \mapsto \mathbf{x}^*.
\end{align}
The proposed DROO algorithm gradually learns such policy function $\pi$ from the experience.

\begin{figure*}
\centering
\includegraphics[width=0.9\textwidth]{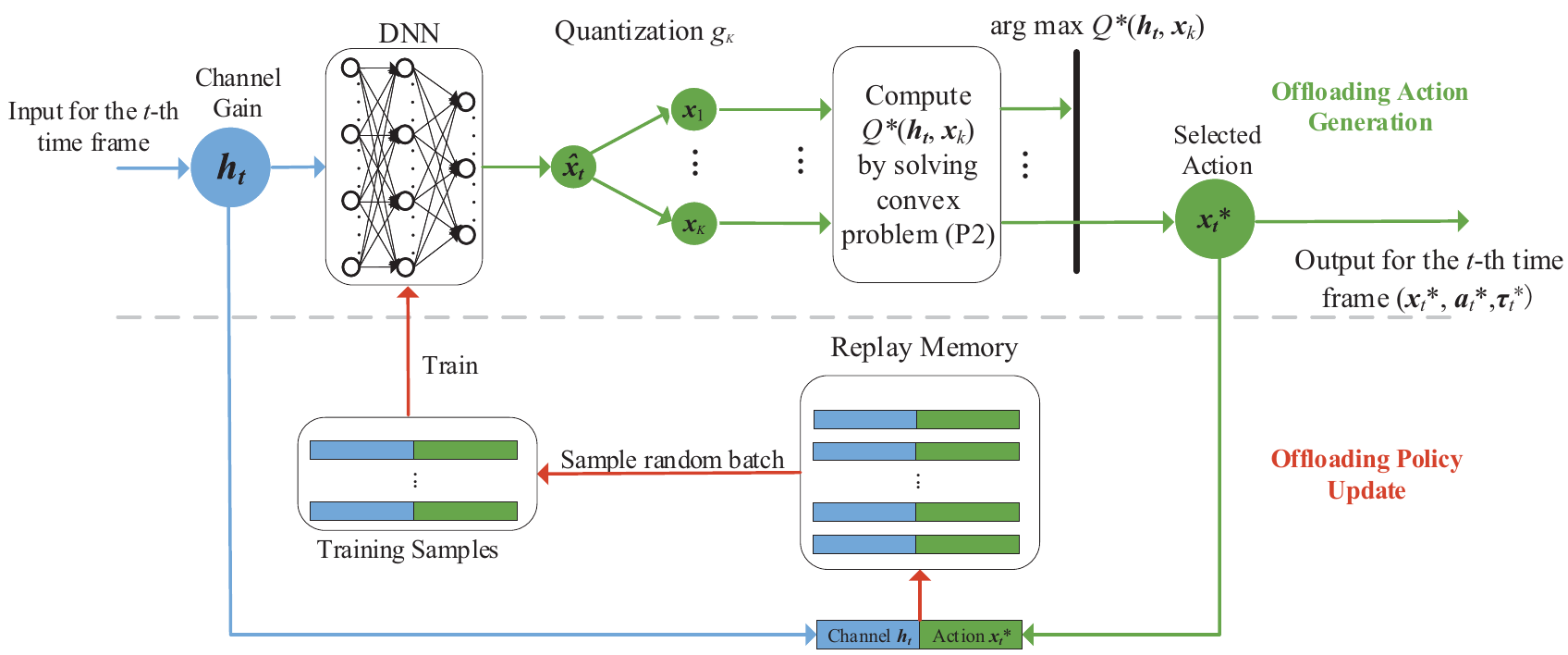}
\caption{The schematics of the proposed DROO algorithm.}
\label{fig:network}
\end{figure*}

\subsection{Algorithm Overview}
The structure of the DROO algorithm is illustrated in Fig.~\ref{fig:network}. It is composed of two alternating stages: offloading action generation and offloading policy update. The generation of the offloading action relies on the use of a DNN, which is characterized by its embedded parameters $\theta$, e.g., the weights that connect the hidden neurons. In the $t$-th time frame, the DNN takes the channel gain $\mathbf{h}_t$ as the input, and outputs a relaxed offloading action $\hat{\mathbf{x}}_t$ (each entry is relaxed to continuous between $0$ and $1$) based on its current offloading policy $\pi_{\theta_{t}}$, parameterized by $\theta_{t}$. The relaxed action is then quantized into $K$ binary offloading actions, among which one best action $\mathbf{x}_t^*$ is selected based on the achievable computation rate as in (P2). {The corresponding $\left\{\mathbf{x}_t^*, a_t^*,\boldsymbol{\tau}_t^*\right\}$ is output as the solution for $\mathbf{h}_t$, which guarantees that all the physical constrains listed in (\ref{equ:p1b})-(\ref{e:binary_x-i}) are satisfied.} The network takes the offloading action $\mathbf{x}_t^*$, receives a reward $Q^*(\mathbf{h}_t,\mathbf{x}_t^*)$, and adds the newly obtained state-action pair $\left(\mathbf{h}_t,\mathbf{x}_t^*\right)$ to the replay memory.

Subsequently, in the policy update stage of the $t$-th time frame, a batch of training samples are drawn from the memory to train the DNN, which accordingly updates its parameter from $\theta_{t}$ to $\theta_{t+1}$ (and equivalently the offloading policy $\pi_{\theta_{t+1}}$). The new offloading policy $\pi_{\theta_{t+1}}$ is used in the next time frame to generate offloading decision $\mathbf{x}_{t+1}^*$ according to the new channel $\mathbf{h}_{t+1}$ observed. Such iterations repeat thereafter as new channel realizations are observed, and the policy $\pi_{\theta_{t}}$ of the DNN is gradually improved. The descriptions of the two stages are detailed in the following subsections.

\subsection{Offloading Action Generation} \label{sec:action_generation}
Suppose that we observe the channel gain realization $\mathbf{h}_t$ in the $t$-th time frame, where $t=1,2,\cdots$. The parameters of the DNN $\theta_t$ are randomly initialized following a zero-mean normal distribution when $t=1$. The DNN first outputs a relaxed computation offloading action $\hat{\mathbf{x}}_t$, represented by a parameterized function $\hat{\mathbf{x}}_t = f_{\theta_t} (\mathbf{h}_t)$, where
\begin{equation}
\hat{\mathbf{x}}_t = \{\hat{x}_{t,i}| \hat{x}_{t,i} \in [0,1], i=1,\cdots,N\}
\end{equation}
and $\hat{x}_{t,i}$ denotes the $i$-th entry of $\hat{\mathbf{x}}_t$.

The well-known universal approximation theorem claims that one hidden layer with enough hidden neurons suffices to approximate any continuous mapping $f$ if a proper activation function is applied at the neurons, e.g., sigmoid, ReLu, and tanh functions \cite{marsland2015ml}. Here, we use ReLU as the activation function in the hidden layers, where the output $y$ and input $v$ of a neuron are related by $y=\max\{v,0\}$. In the output layer, we use a sigmoid activation function, i.e., $y= 1/\left(1+e^{-v}\right)$, such that the relaxed offloading action satisfies $\hat{x}_{t,i} \in (0,1)$.

Then, we quantize $\hat{\mathbf{x}}_t$ to obtain $K$ binary offloading actions, where $K$ is a design parameter. The quantization function, $g_K$, is defined as
\begin{align}
  g_K:\hat{\mathbf{x}}_t \mapsto \{\mathbf{x}_k \mid \mathbf{x}_k \in \{0,1\}^N, k =1,\cdots, K\}.
\end{align}
In general, $K$ can be any integer within $[1, 2^N]$ ($N$ is the number of WDs), where a larger $K$ results in better solution quality and higher computational complexity, and vice versa. To balance the performance and complexity, we propose an \textit{order-preserving quantization} method, where the value of $K$ could be set from $1$ to $(N+1)$. The basic idea is to preserve the ordering during quantization. That is, for each quantized action $\mathbf{x}_k$, $x_{k,i} \geq x_{k,j}$ should hold if $\hat{x}_{t,i}\geq \hat{x}_{t,j}$ for all $i,j \in \left\{1,\cdots, N\right\}$. Specifically, for a given $1\leq K\leq N+1$, the set of $K$ quantized actions $\{\mathbf{x}_k\}$ is generated from the relaxed action $\hat{\mathbf{x}}_t$ as follows:
\begin{enumerate}
  \item The first binary offloading decision $\mathbf{x}_1$ is obtained as
\begin{align}
\label{24}
x_{1,i} =  \begin{cases}
                          1 & \hat{x}_{t,i} > 0.5, \\
                          0 & \hat{x}_{t,i} \leq 0.5,
                   \end{cases}
\end{align}
for $i=1, \cdots, N$.
  \item To generate the remaining $K-1$ actions, we first order the entries of $\hat{\mathbf{x}}_t$ with respective to their distances to $0.5$, denoted by $ \lvert \hat{x}_{t,(1)}-0.5 \rvert \leq \lvert \hat{x}_{t,(2)}-0.5 \rvert \leq \dots \leq \lvert \hat{x}_{t,(i)}-0.5 \rvert \dots \leq \lvert \hat{x}_{t,(N)}-0.5 \rvert$, where $\hat{x}_{t,(i)}$ is the $i$-th order statistic of $\hat{\mathbf{x}}_t$. Then, the $k$-th offloading decision $\mathbf{x}_k$, where $k = 2,\cdots, K$, is calculated based on $\hat{x}_{t,(k-1)}$ as
\begin{align}
\label{23}
x_{k,i} =  \begin{cases}
                          1 & \hat{x}_{t,i} > \hat{x}_{t,(k-1)}, \\
                          1 & \hat{x}_{t,i} = \hat{x}_{t,(k-1)}\ \textrm{and}\ \hat{x}_{t,(k-1)} \leq 0.5, \\
                          0 & \hat{x}_{t,i} = \hat{x}_{t,(k-1)}\ \textrm{and}\ \hat{x}_{t,(k-1)} > 0.5, \\
                          0 & \hat{x}_{t,i} < \hat{x}_{t,(k-1)},
                   \end{cases}
\end{align}
for $i=1, \cdots, N$.
\end{enumerate}
Because there are in total $N$ order statistic of $\hat{\mathbf{x}}_t$, while each can be used to generate one quantized action from (\ref{23}), the above order-preserving quantization method in (\ref{24}) and (\ref{23}) generates at most $(N+1)$ quantized actions, i.e., $K\leq N+1$. In general, setting a large $K$ (e.g., $K=N$) leads to better computation rate performance at the cost of higher complexity. However, as we will show later in Section~\ref{sec:adaptive}, it is not only inefficient but also unnecessary to generate a large number of quantized actions in each time frame. Instead, setting a small $K$ (even close to $1$) suffices to achieve good computation rate performance and low complexity after sufficiently long training period.

We use an example to illustrate the above order-preserving quantization method. Suppose that $\hat{\mathbf{x}}_t$ = [0.2, 0.4, 0.7, 0.9] and $K=4$. {The corresponding order statistics of $\hat{\mathbf{x}}_t$ are $  \hat{x}_{t,(1)} = 0.4 $, $  \hat{x}_{t,(2)} = 0.7 $, $  \hat{x}_{t,(3)} = 0.2 $, and $  \hat{x}_{t,(4)} = 0.9 $.} Therefore, the $4$ offloading actions generated from the above quantization method are $\mathbf{x}_1$ = [0, 0, 1, 1], $\mathbf{x}_2$ = [0, 1, 1, 1], $\mathbf{x}_3$ = [0, 0, 0, 1], and $\mathbf{x}_4$ = [1, 1, 1, 1]. In comparison, when the conventional KNN method is used, the obtained actions are $\mathbf{x}_1$ = [0, 0, 1, 1], $\mathbf{x}_2 $= [0, 1, 1, 1], $\mathbf{x}_3$ =  [0, 0, 0, 1], and $\mathbf{x}_4$ = [0, 1, 0, 1].

Compared to the KNN method where the quantized solutions are closely placed around $\hat{x}$, the offloading actions produced by the order-preserving quantization method are separated by a larger distance. Intuitively, this creates higher diversity in the candidate action set, thus increasing the chance of finding a local maximum around $\hat{\mathbf{x}}_t$. In Section~\ref{sec:DROO_performance}, we show that the proposed order-preserving quantization method achieves better convergence performance than KNN method.

Recall that each candidate action $\mathbf{x}_k$ can achieve $Q^*(\mathbf{h}_t,\mathbf{x}_k)$ computation rate by solving (P2). Therefore, the best offloading action $\mathbf{x}^*_t$ at the $t$-th time frame is chosen as
\begin{align}
  \mathbf{x}^*_t = \arg \underset{\mathbf{x}_i\in \left\{\mathbf{x}_k\right\}}{ \max} \ Q^*(\mathbf{h}_t,\mathbf{x}_i).
  \label{e:x_star}
\end{align}
Note that the $K$-times evaluation of $Q^*(\mathbf{h}_t,\mathbf{x}_k)$ can be processed in parallel to speed up the computation of (\ref{e:x_star}). {Then, the network outputs the offloading action $\mathbf{x}^*_t$ along with its corresponding optimal resource allocation $ (\mathbf{\tau}^*_t, a^*_t) $}.

\subsection{Offloading Policy Update}
The offloading solution obtained in (\ref{e:x_star}) will be used to update the offloading policy of the DNN. Specifically, we maintain an initially empty memory of limited capacity. At the $t$-th time frame, a new training data sample $(\mathbf{h}_t,\mathbf{x}^*_t)$ is added to the memory. When the memory is full, the newly generated data sample replaces the oldest one.

We use the experience replay technique \cite{lin1993reinforcement, mnih2015nature} to train the DNN using the stored data samples. In the $t$-th time frame, we randomly select a batch of training data samples $\{(\mathbf{h}_{\tau},\mathbf{x}^*_{\tau})\mid {\tau} \in \mathcal{T}_t\}$ from the memory, characterized by a set of time indices $\mathcal{T}_t$. The parameters $\theta_t$ of the DNN are updated by applying the Adam algorithm \cite{kingma2014adam} to reduce the averaged cross-entropy loss, as
\begin{equation*}
\small
\begin{aligned}
  &L(\theta_t) =\\
   &-\frac{1}{|\mathcal{T}_t|}{\sum}_{\tau \in \mathcal{T}_t}\Big( {(\mathbf{x}^*_{\tau})}^{\intercal} \log f_{\theta_t}(\mathbf{h}_{\tau}) + (1-{\mathbf{x}^*_{\tau}})^{\intercal}\log \big(1-f_{\theta_t}(\mathbf{h}_{\tau})\big)\Big),
   \end{aligned}
\end{equation*}
where $|\mathcal{T}_t|$ denotes the size of $\mathcal{T}_t$, the superscript $\intercal$ denotes the transpose operator, and the log function denotes the element-wise logarithm operation of a vector. The detailed update procedure of the Adam algorithm is omitted here for brevity. In practice, we train the DNN every $\delta$ time frames after collecting sufficient number of new data samples. The experience replay technique used in our framework has several advantages. First, the batch update has a reduced complexity than using the entire set of data samples. Second, the reuse of historical data reduces the variance of $\theta_t$ during the iterative update. Third, the random sampling fastens the convergence by reducing the correlation in the training samples.

Overall, the DNN iteratively learns from the best state-action pairs $\left(\mathbf{h}_t, \mathbf{x}^*_t\right)$'s and generates better offloading decisions output as the time progresses. Meanwhile, with the finite memory space constraint, the DNN only learns from the most recent data samples generated by the most recent (and more refined) offloading policies. This closed-loop reinforcement learning mechanism constantly improves its offloading policy until convergence. We provide the pseudo-code of the DROO algorithm in Algorithm~1.

\begin{algorithm}
 \SetAlgoLined
 \SetKwData{Left}{left}\SetKwData{This}{this}\SetKwData{Up}{up}
 \SetKwRepeat{doWhile}{do}{while}
 \SetKwFunction{Union}{Union}\SetKwFunction{FindCompress}{FindCompress}
 \SetKwInOut{Input}{input}\SetKwInOut{Output}{output}
  \Input{Wireless channel gain $\mathbf{h}_t$ at each time frame $t$, the number of quantized actions $K$}
 \Output{Offloading action $\mathbf{x}^*_t$, and the corresponding optimal resource allocation for each time frame $t$;}
 Initialize the DNN with random parameters $\theta_{1}$ and empty memory\;
 Set iteration number $M$ and the training interval $\delta$\;
 \For{$t=1,2,\dots,M$}{
 Generate a relaxed offloading action $\hat{\mathbf{x}}_t = f_{\theta_t}(\mathbf{h}_t)$\;
 Quantize $\hat{\mathbf{x}}_t$ into $K$ binary actions $\{\mathbf{x}_k\} = g_K(\hat{\mathbf{x}}_t)$\;
 Compute $Q^*(\mathbf{h}_t,\mathbf{x}_k)$ for all $\{\mathbf{x}_k \}$ by solving (P2)\; \label{step:compute_Q}
 Select the best action $\mathbf{x}^*_{t}=\arg \underset{\{\mathbf{x}_k\}}{\max}\ Q^*(\mathbf{h}_t,\mathbf{x}_k)$\;
 Update the memory by adding $(\mathbf{h}_t,\mathbf{x}^*_t)$\;
 \If{$t \bmod \delta = 0$}{
   Uniformly sample a batch of data set $\{(\mathbf{h}_{\tau}, \mathbf{x}^*_{\tau})\mid \tau \in \mathcal{T}_t\}$ from the memory\;
   Train the DNN with $\{(\mathbf{h}_{\tau}, \mathbf{x}^*_{\tau})\mid \tau \in \mathcal{T}_t\}$ and update $\theta_{t}$ using the Adam algorithm\;
 }
 }
 \caption{An online DROO algorithm to solve the offloading decision problem.}
\end{algorithm}

\subsection{Adaptive Setting of $K$} \label{sec:adaptive}
Compared to the conventional optimization algorithms, the DROO algorithm has the advantage in removing the need of solving hard MIP problems, and thus has the potential to significantly reduce the complexity. The major computational complexity of the DROO algorithm comes from solving (P2) $K$ times in each time frame to select the best offloading action. Evidently, a larger $K$ (e.g., $K=N$) in general leads to a better offloading decision in each time frame and accordingly a better offloading policy in the long term. Therefore, there exists a fundamental performance-complexity tradeoff in setting the value of $K$.

\begin{figure}
\centering
  \begin{center}
    \includegraphics[width=0.5\textwidth]{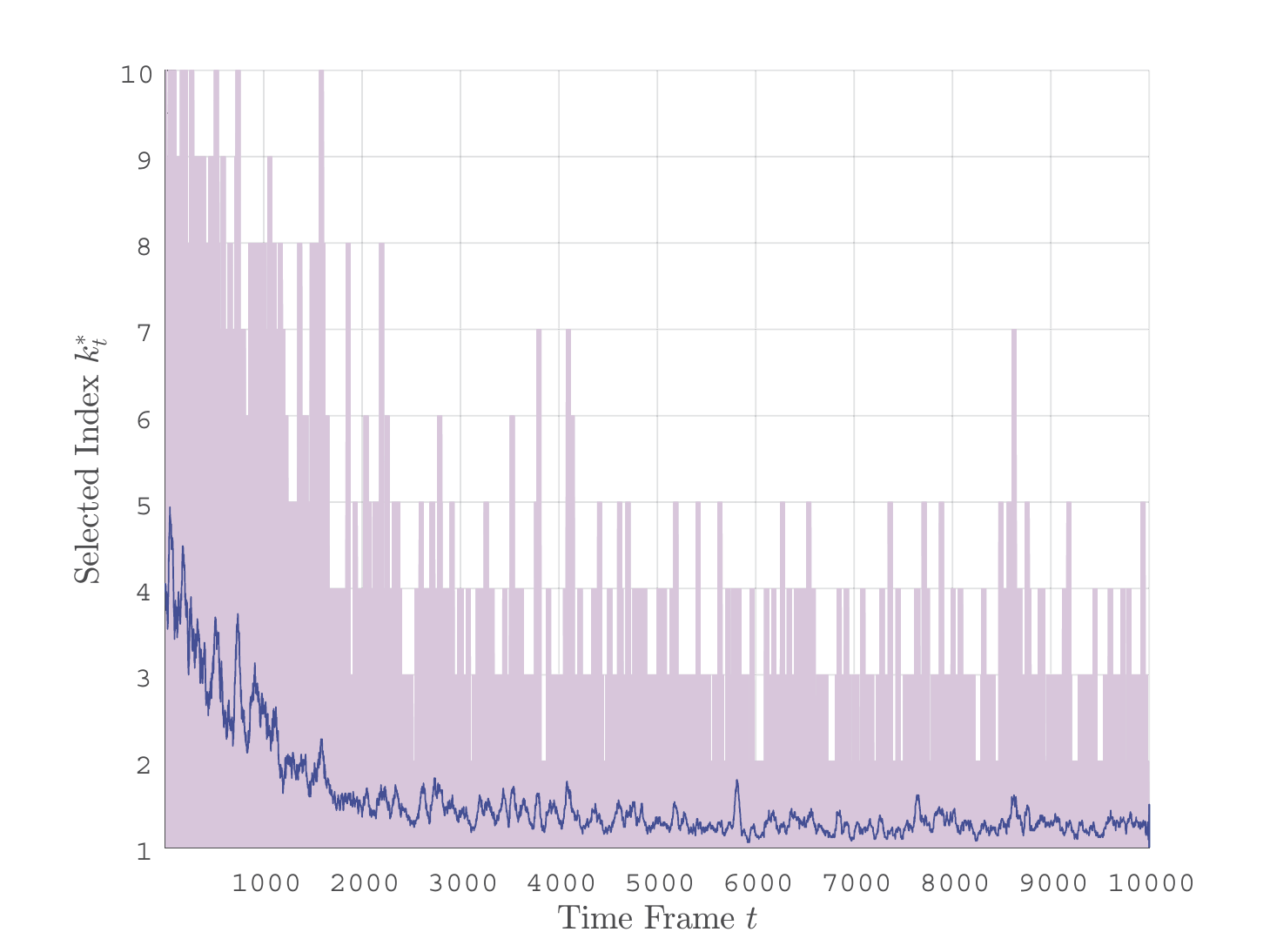}
  \end{center}
  \caption{The index $k^*_t$ of the best offloading actions $\mathbf{x}^*_t$ for DROO algorithm when the number of WDs is $N=10$ and $K=N$. The detailed simulation setups are presented in Section~\ref{sec:result}.}
  \label{fig:ordered_index}
\end{figure}

In this subsection, we propose an adaptive procedure to automatically adjust the number of quantized actions generated by the order-preserving quantization method. We argue that using a large and fixed $K$ is not only computationally inefficient but also unnecessary in terms of computation rate performance. To see this, consider a wireless powered MEC network with $N=10$ WDs. We apply the DROO algorithm with a fixed $K=10$ and plot in Fig.~\ref{fig:ordered_index} the index of the best action $\mathbf{x}^*_t$ calculated from (\ref{e:x_star}) over time, denoted as $k^*_t$. For instance, $k^*_t = 2$ indicates that the best action in the $t$-th time frame is ranked the second among the $K$ ordered quantized actions. In the figure, the curve is plotted as the $50$-time-frames rolling average of $k^*_t$ and the light shadow region is the upper and lower bounds of $k^*_t$ in the past $50$ time frames. Apparently, most of the selected indices $k^*_t$ are no larger than $5$ when $t \geq 5000$. This indicates that those generated offloading actions $\mathbf{x}_{k}$ with $k >5$ are redundant. In other words, we can gradually reduce $K$ during the learning process to speed up the algorithm without compromising the performance.

Inspired by the results in Fig.~\ref{fig:ordered_index}, we propose an adaptive method for setting $K$. We denote $K_t$ as the number of binary offloading actions generated by the quantization function at the $t$-th time frame. We set $K_1 = N$ initially and update $K_t$ every $\Delta$ time frames, where $\Delta$ is referred to as the updating interval for $K$. Upon an update time frame, $K_t$ is set as $1$ plus the largest $k^*_t$ observed in the past $\Delta$ time frames. The reason for the additional $1$ is to allow $K_t$ to increase during the iterations. Mathematically, $K_t$ is calculated as
\begin{align*}
  K_{t }= \begin{cases}
  N,  & t = 1,\\
  \min\left(\max \left(k_{t - 1}^*,\cdots, k_{t - \Delta}^* \right)+1,N\right),  & t \bmod \Delta = 0, \\
                                          K_{t-1 },  & \textrm{otherwise}, \\
                    \end{cases}
\end{align*}
for $t \geq 1$. For an extreme case with $\Delta =1$, $K_t$ updates in each time frame. Meanwhile, when $\Delta \rightarrow \infty$, $K_t$ never updates such that it is equivalent to setting a constant $K=N$. In Section~\ref{sec:adaptive_performance}, we numerically show that setting a proper $\Delta$ can effectively speed up the learning process without compromising the computation rate performance.

\section{Numerical Results}\label{sec:result}
In this section, we use simulations to evaluate the performance of the proposed DROO algorithm. In all simulations, we use the parameters of Powercast TX91501-3W with $P=3$ Watts for the energy transmitter at the AP, and those of P2110 Powerharvester for the energy receiver at each WD.\footnote{See detailed product specifications at \url{http://www.powercastco.com.}} The energy harvesting efficiency $\mu= 0.51$. The distance from the $i$-th WD to the AP, denoted by $d_i$, is uniformly distributed in the range of (2.5, 5.2) meters, $i=1,\cdots,N$. Due to the page limit, the exact values of $d_i$'s are omitted. The average channel gain $\bar{h}_i$ follows the free-space path loss model $\bar{h}_i = A_d\left(\frac{3\cdot10^8}{4\pi f_c d_i}\right)^{d_e}$, where $A_d = 4.11$ denotes the antenna gain, $f_c =915$ MHz denotes the carrier frequency, and $d_e = 2.8$ denotes the path loss exponent. The time-varying wireless channel gain of the $N$ WDs at time frame $t$, denoted by $\mathbf{h}_t = \left[h^t_1,h^t_2,\cdots, h^t_N\right]$, {is generated from a Rayleigh fading channel model} as $h^t_i = \bar{h}_i \alpha^t_{i}$. Here $\alpha^t_{i}$ is the independent random channel fading factor following an exponential distribution with unit mean. Without loss of generality, the channel gains are assumed to remain the same within one time frame and vary independently from one time frame to another. We assume equal computing efficiency $k_i = 10^{-26}$, $i=1,\cdots,N$, and $\phi=100$ for all the WDs \cite{wang2016toc}. The data offloading bandwidth $B=2$ MHz, receiver noise power $N_0=10^{-10}$, and $v_u = 1.1$. Without loss of generality, we set $T=1$ and the $w_i=1$ if $i$ is an odd number and $w_i=1.5$ otherwise. All the simulations are performed on a desktop with an Intel Core i5-4570 3.2 GHz CPU and 12 GB memory.

We simply consider a fully connected DNN consisting of one input layer, two hidden layers, and one output layer in the proposed DROO algorithm, where the first and second hidden layers have $120$ and $80$ hidden neurons, respectively. Note that the DNN can be replaced by other structures with different number of hidden layers and neurons, or even other types of neural networks to fit the specific learning problem, such as convolutional neural network (CNN) or recurrent neural network (RNN) \cite{Goodfellow-et-al-2016}. In this paper, we find that a simple two-layer perceptron suffices to achieve satisfactory convergence performance, while better convergence performance is expected by further optimizing the DNN parameters. We implement the DROO algorithm in Python with TensorFlow 1.0 and set training interval $\delta=10$, training batch size $|\mathcal{T}| = 128$, memory size as 1024, and learning rate for Adam optimizer as 0.01. {The source code is available at \url{https://github.com/revenol/DROO}}.

\subsection{Convergence Performance} \label{sec:DROO_performance}
We first consider a wireless powered MEC network with $N=10$ WDs. Here, we define the normalized computation rate $\hat{Q}( \mathbf{h}, \mathbf{x})\in [0,1]$, as
\begin{align}
\hat{Q} (\mathbf{h}, \mathbf{x}) = \frac{Q^*(\mathbf{h},\mathbf{x})}{\max_{\mathbf{x}' \in \{0,1\}^N}Q^*(\mathbf{h},\mathbf{x}')},
\label{e:Q_N}
\end{align}
where the optimal solution in the denominator is obtained by enumerating all the $2^N$ offloading actions.

\begin{figure}
    \centering
    \begin{center}
        \includegraphics[width=0.47\textwidth]{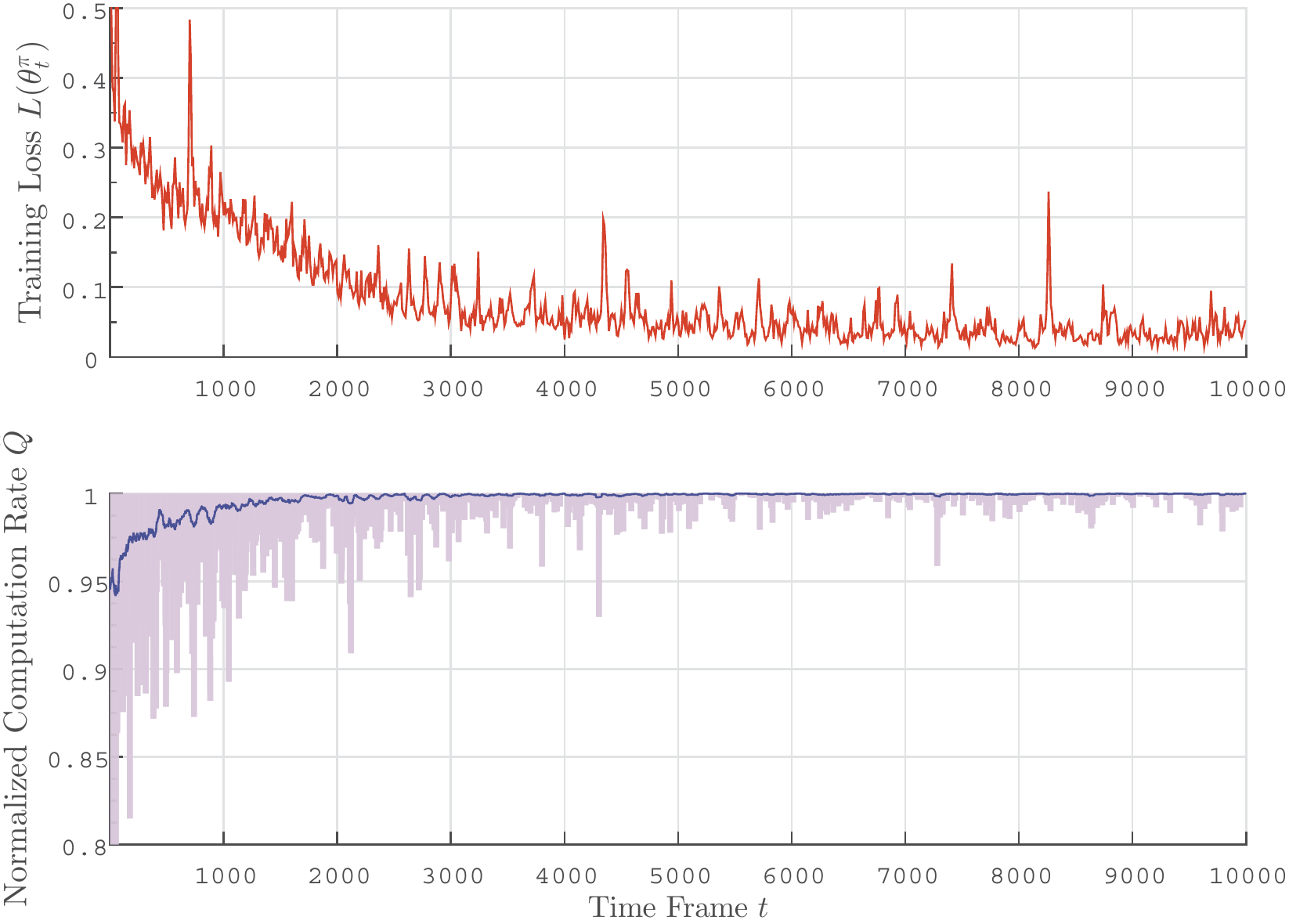}
    \end{center}
    \caption{Normalized computation rates and training losses for DROO algorithm under fading channels when $N=10$ and $K=10$.}
    \label{fig:example10user9997}
\end{figure}

In Fig.~\ref{fig:example10user9997}, we plot the training loss $L(\theta_t)$ of the DNN and the normalized computation rate $\hat{Q}$. Here, we set a fixed $K=N$. In the figure below, the blue curve denotes the moving average of $\hat{Q}$ over the last $50$ time frames, and the light blue shadow denotes the maximum and minimum of $\hat{Q}$ in the last $50$ frames. We see that the moving average $\hat{Q}$ of DROO gradually converges to the optimal solution when $t$ is large. Specifically, the achieved average $\hat{Q}$ exceeds 0.98 at an early stage when $t>400$ and the variance gradually decreases to zero as $t$ becomes larger, e.g., when $t>3,000$. Meanwhile, in the figure above, the training loss $L(\theta_t)$ gradually decreases and stabilizes at around 0.04, {whose fluctuation is mainly due to the random sampling of training data}.

\begin{figure}
    \centering
    \begin{center}
        \includegraphics[width=0.47\textwidth]{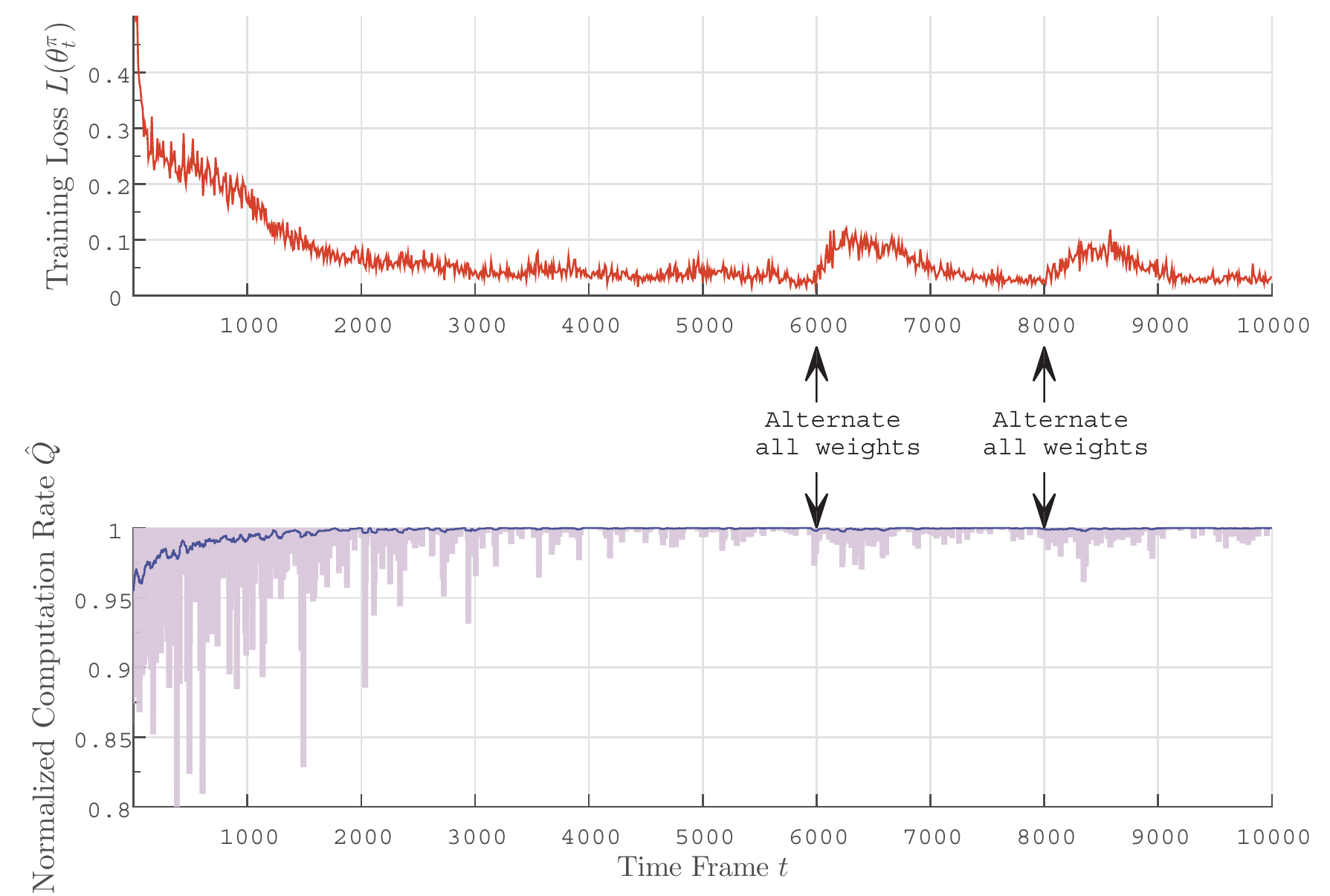}
    \end{center}
    \caption{Normalized computation rates and training losses for DROO algorithm with alternating-weight WDs when $N=10$ and $K=10$.}
    \label{fig:priority}
\end{figure}
{In Fig.~\ref{fig:priority}, we evaluate DROO for MEC networks with alternating-weight WDs. We evaluate the worst case by alternating the weights of all WDs between 1 and 1.5 at the same time, specifically, at $t=6,000$ and $t=8,000$. The training loss sharply increases after the weights alternated and gradually decreases and stabilizes after training for 1,000 time frames, which means that DROO automatically updates its offloading decision policy and converges to the new optimal solution. Meanwhile, as shown in Fig.~\ref{fig:priority}, the minimum of  $\hat{Q}$ is greater than 0.95 and the moving average of $\hat{Q}$ is always greater than 0.99 for $t>6,000$.}

\begin{figure}
    \centering
    \begin{center}
        \includegraphics[width=0.47\textwidth]{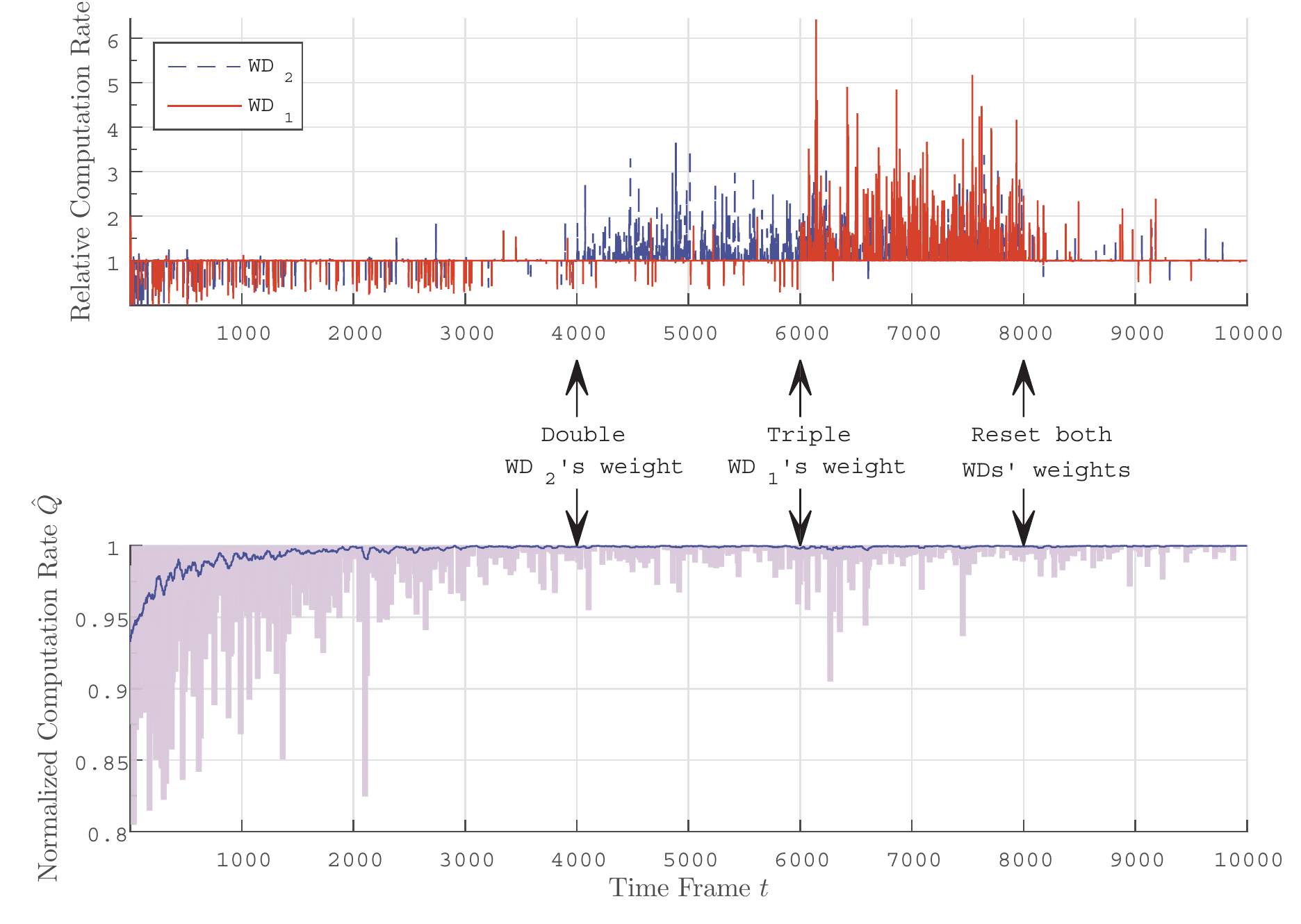}
    \end{center}
    \caption{Computation rates for DROO algorithm with temporarily new weights when $N=10$ and $K=10$.}
    \label{fig:doublepriority}
\end{figure}

{In Fig.~\ref{fig:doublepriority}, we evaluate the ability of DROO in supporting WDs' temporarily critical computation demand. Suppose that $WD_1$ and $WD_2$ have a temporary surge of commutation demands. We double $WD_2$'s weight from 1.5 to 3 at time frame $t=4,000$, triple $WD_1$'s weight from 1 to 3 at $t=6,000$, and reset both of their weights to the original values at $t=8,000$. In the top sub-figure in Fig.~\ref{fig:doublepriority}, we plot the relative computation rates for both WDs, where each WD's computation rate is normalized  against that achieved under the optimal offloading actions with their original weights. In the first 3,000 time frames, DROO gradually converges and the corresponding relative computation rates for both WDs are lower than the baseline at most of the time frames. During time frames $4,000<t<8,000$, $WD_2$'s weight is doubled. Its computation rate significantly improves over the baseline, where at some time frames the improvement can be as high as 2 to 3 times of the baseline. Similar rate improvement is also observed for $WD_1$ when its weight is tripled between  $6,000<t<8,000$. In addition, their computation rates gradually converge to the baseline when their weights are reset to the original value after $t =8,000$. On average, $WD_1$ and $WD_2$ have experienced  26\% and 12\% higher computation rate, respectively, during their periods with increased weights. In the bottom sub-figure in Fig.~\ref{fig:doublepriority}, we plot the normalized computation rate performance of DROO, which shows that the algorithm can quickly adapt itself to the temporary demand variation of users. The results in Fig.~\ref{fig:doublepriority} have verified the ability of the propose DROO framework in supporting temporarily critical service quality requirements.}

\begin{figure}
    \centering
    \begin{center}
        \includegraphics[width=0.47\textwidth]{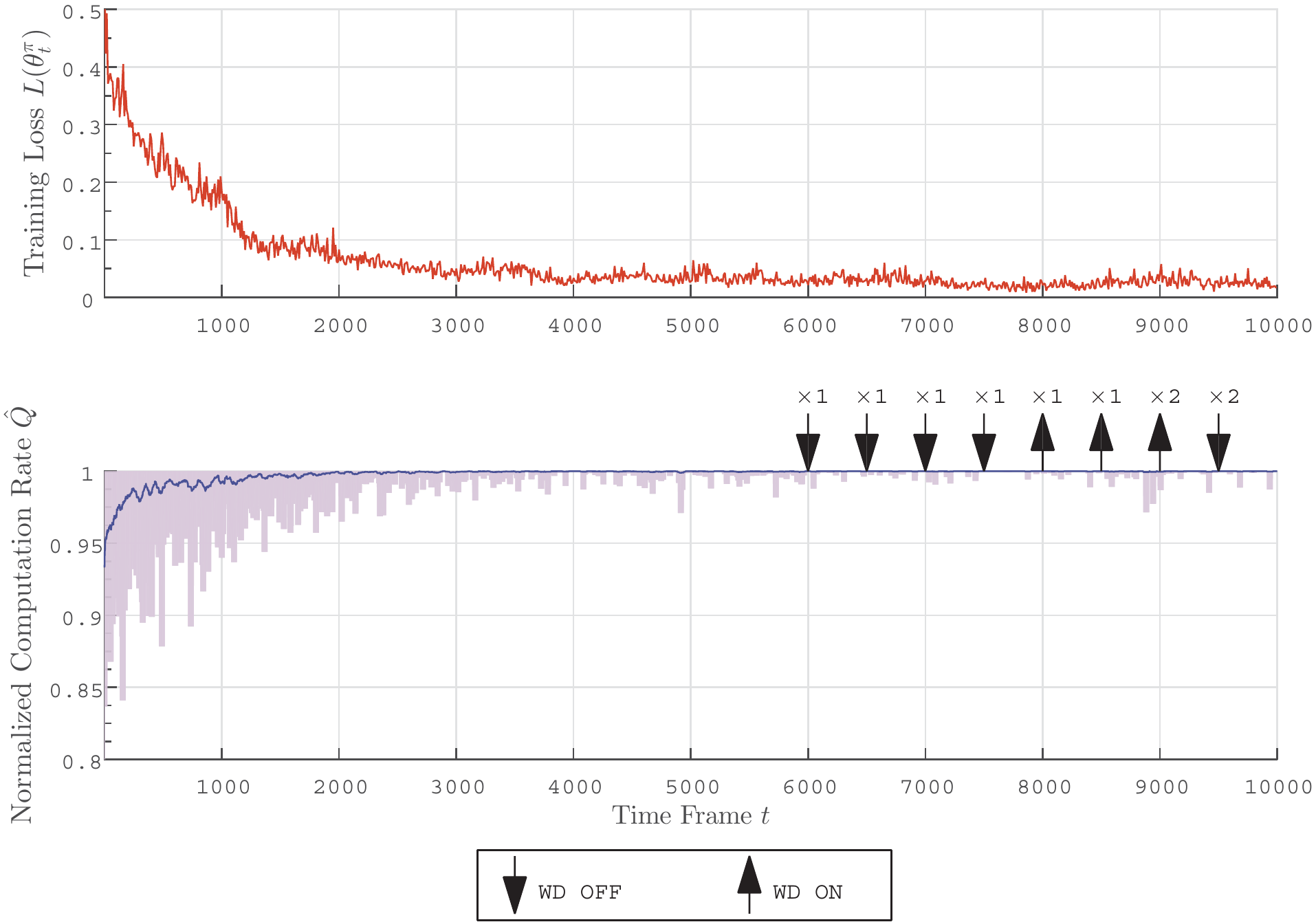}
    \end{center}
    \caption{Normalized computation rates and training losses for DROO algorithm with ON-OFF WDs when $N=10$ and $K=10$.}
    \label{fig:on_off_10}
\end{figure}
{In Fig.~\ref{fig:on_off_10}, we evaluate DROO for MEC networks where WDs can be occasionally turned off/on. After DROO converges, we randomly turn off on one WD at each time frame $t=6,000, 6,500, 7,000, 7,500$, and then turn them on at time frames $t=8,000, 8,500, 9,000$. At time frame $t=9,500$, we randomly turn off two WDs, resulting an MEC network with 8 active WDs. {Since the number of neurons in the input layer of DNN is fixed as $N=10$, we set the input channel gains $h$ for the inactive WDs as 0 to exclude them from the resource allocation optimization with respect to (P2)}. We numerically study the performance of this modified DROO in Fig.~\ref{fig:on_off_10}. Note that, when evaluating the normalized computation rate $\hat{Q}$ via equation (\ref{e:Q_N}), the denominator is re-computed when one WD is turned off/on. For example, when there are 8 active WDs in the MEC network, the denominator is obtained by enumerating all the $2^8$ offloading actions.  As shown in Fig.~\ref{fig:on_off_10}, the training loss $L(\theta_t)$ increases little after WDs are turned off/on, and the moving average of the resulting $\hat{Q}$ is always greater than 0.99.}

\begin{figure*}
    \centering
    \begin{center}
        \includegraphics[width=0.85\textwidth]{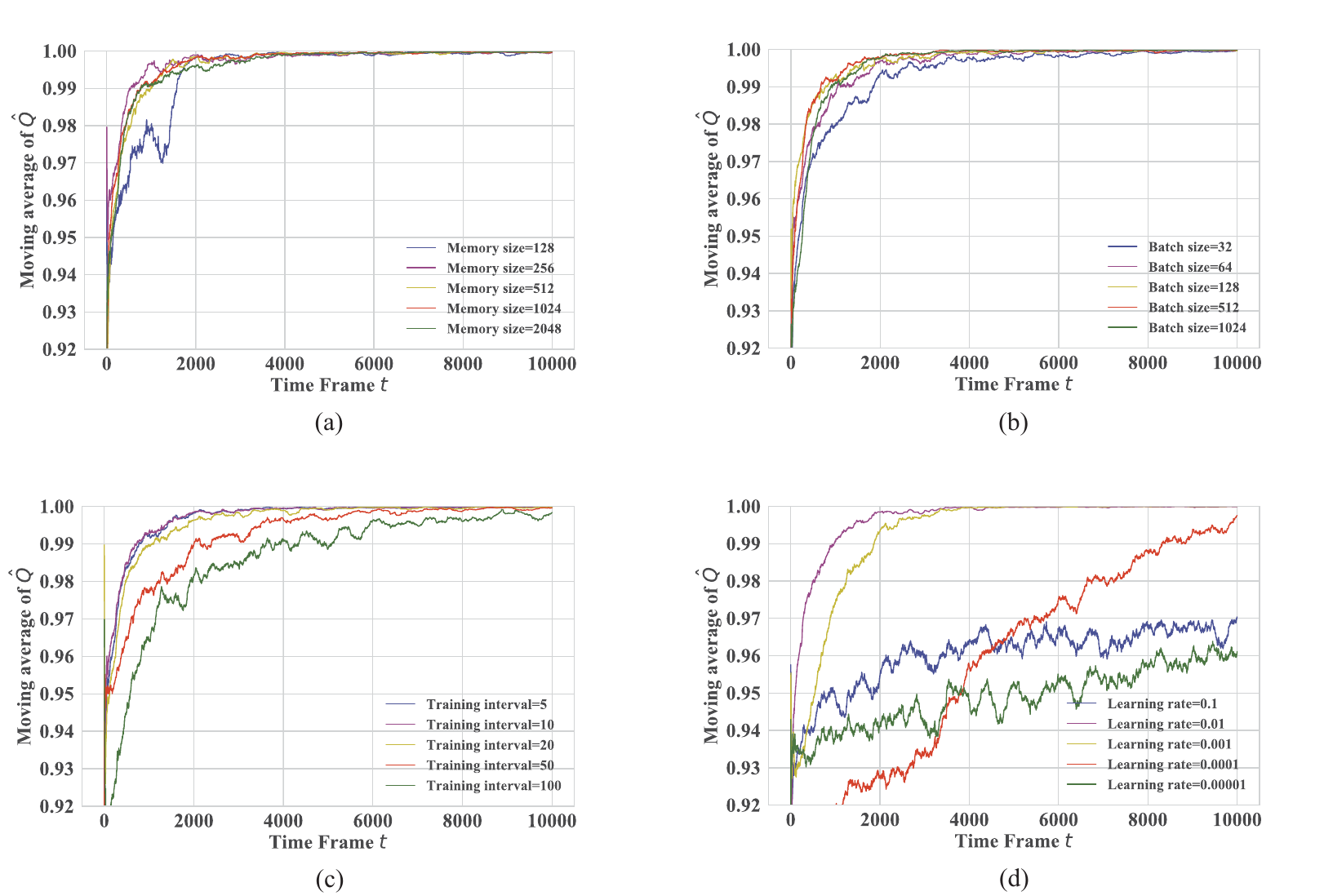}
    \end{center}
    \caption{Moving average of $\hat{Q}$ under different algorithm parameters when $N=10$: (a) memory size ; (b) training batch size; (c) training interval; (d) learning rate.}
    \label{fig:para}
\end{figure*}
In Fig.~\ref{fig:para}, we further study the effect of different algorithm parameters on the convergence performance of DROO, including different memory sizes, batch sizes, training intervals, and learning rates. In Fig.~\ref{fig:para}(a), a small memory (=128) causes larger fluctuations on the convergence performance, while a large memory (=2048) requires more training data to converge to optimal, as $\hat{Q} = 1$. In the following simulations, we choose the memory size as 1024. For each training procedure, we randomly sample a batch of data samples from the memory to improve the DNN. Hence, the batch size must be no more than the memory size 1024. As shown in Fig.~\ref{fig:para}(b), a small batch size (=32) does not take advantage of all training data stored in the memory, while a large batch size (=1024) frequently uses the ``old'' training data and degrades the convergence performance. Furthermore, a large batch size consumes more time for training. As a trade-off between convergence speed and computation time, we set the training batch size $|\mathcal{T}| = 128$ in the following simulations. In Fig.~\ref{fig:para}(c), we investigate the convergence of DROO under different training intervals $\delta$. DROO converges faster with shorter training interval, and thus more frequent policy update. However, numerical results show that it is unnecessary to train and update the DNN too frequently. Hence, we set the training interval $\delta=10$ to speed up the convergence of DROO. In Fig.~\ref{fig:para}(d), we study the impact of the learning rate in Adam optimizer \cite{kingma2014adam} to the convergence performance. We notice that either a too small or a too large learning rate causes the algorithm to converge to a local optimum. In the following simulations, we set the learning rate as 0.01.

\begin{figure}
    \centering
    \begin{center}
        \includegraphics[width=0.45\textwidth]{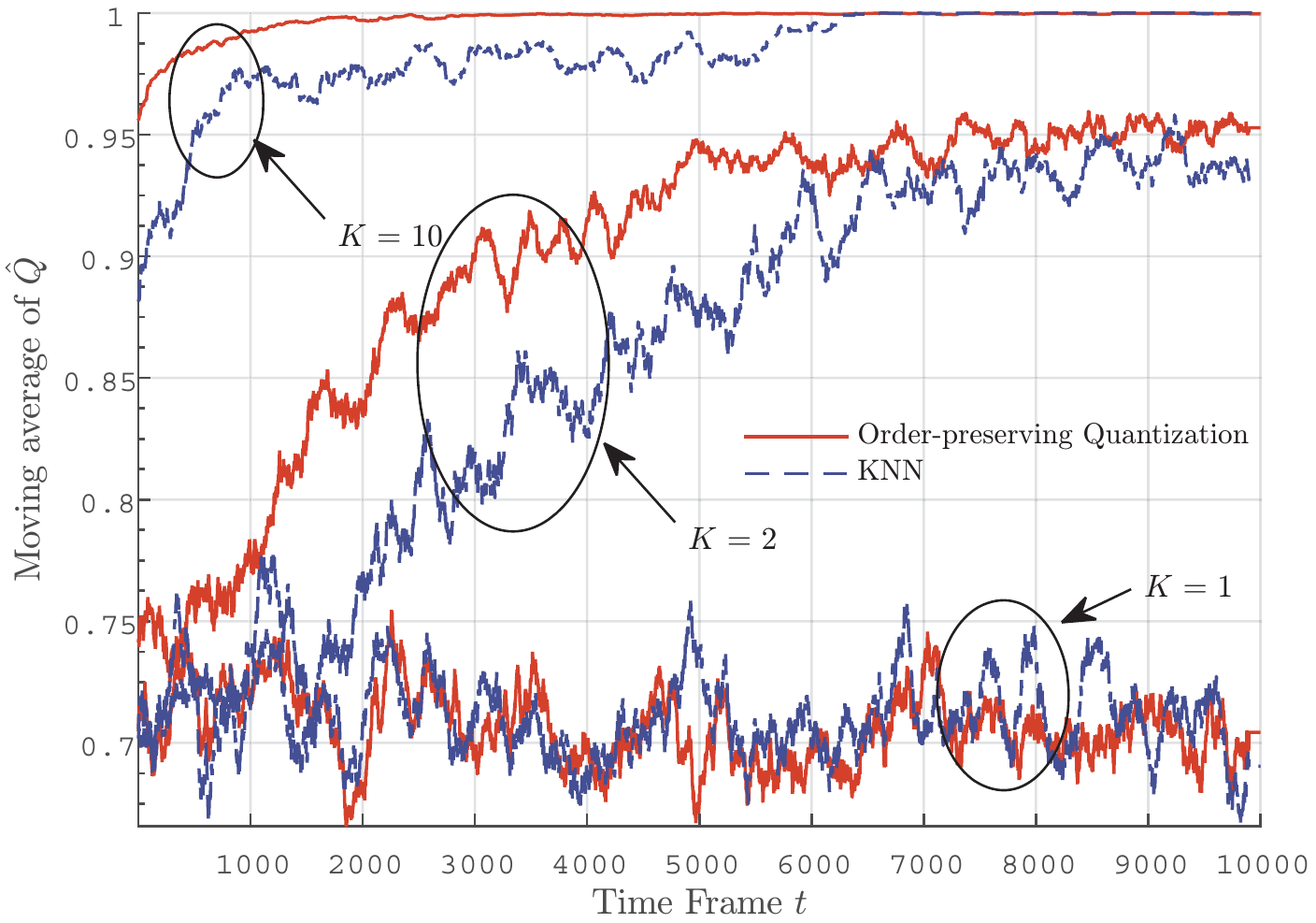}
    \end{center}
    \caption{Moving average of $\hat{Q}$ under different quantization functions and $K$ when $N=10$.}
    \label{fig:knm_vs_knn}
\end{figure}

In Fig.~\ref{fig:knm_vs_knn}, we compare the performance of two quantization methods: the proposed order-preserving quantization and the conventional KNN quantization method under different $K$. In particular, we plot the the moving average of $\hat{Q}$ over a window of $200$ time frames. When $K = N$, both methods converge to the optimal offloading actions, i.e., the moving average of $\hat{Q}$ approaches $1$. However, they both achieve suboptimal offloading actions when $K$ is small. For instance, when $K=2$, the order-preserving quantization method and KNN both only converge to around $0.95$. Nonetheless, we can observe that when $K\geq 2$, the order-preserving quantization method converges faster than the KNN method. Intuitively, this is because the order-preserving quantization method offers a larger diversity in the candidate actions than the KNN method. Therefore, the training of DNN requires exploring fewer offloading actions before convergence. Notice that the DROO algorithm does not converge for both quantization methods when $K=1$. This is because the DNN cannot improve its offloading policy when action selection is absent.

{The simulation results in this subsection show that the proposed DROO framework can quickly converge to the optimal offloading policy}, especially when the proposed order-preserving action quantization method is used.

\subsection{Impact of Updating Intervals $\Delta$} \label{sec:adaptive_performance}
\begin{figure}
    \centering
    \begin{center}
        \includegraphics[width=0.47\textwidth]{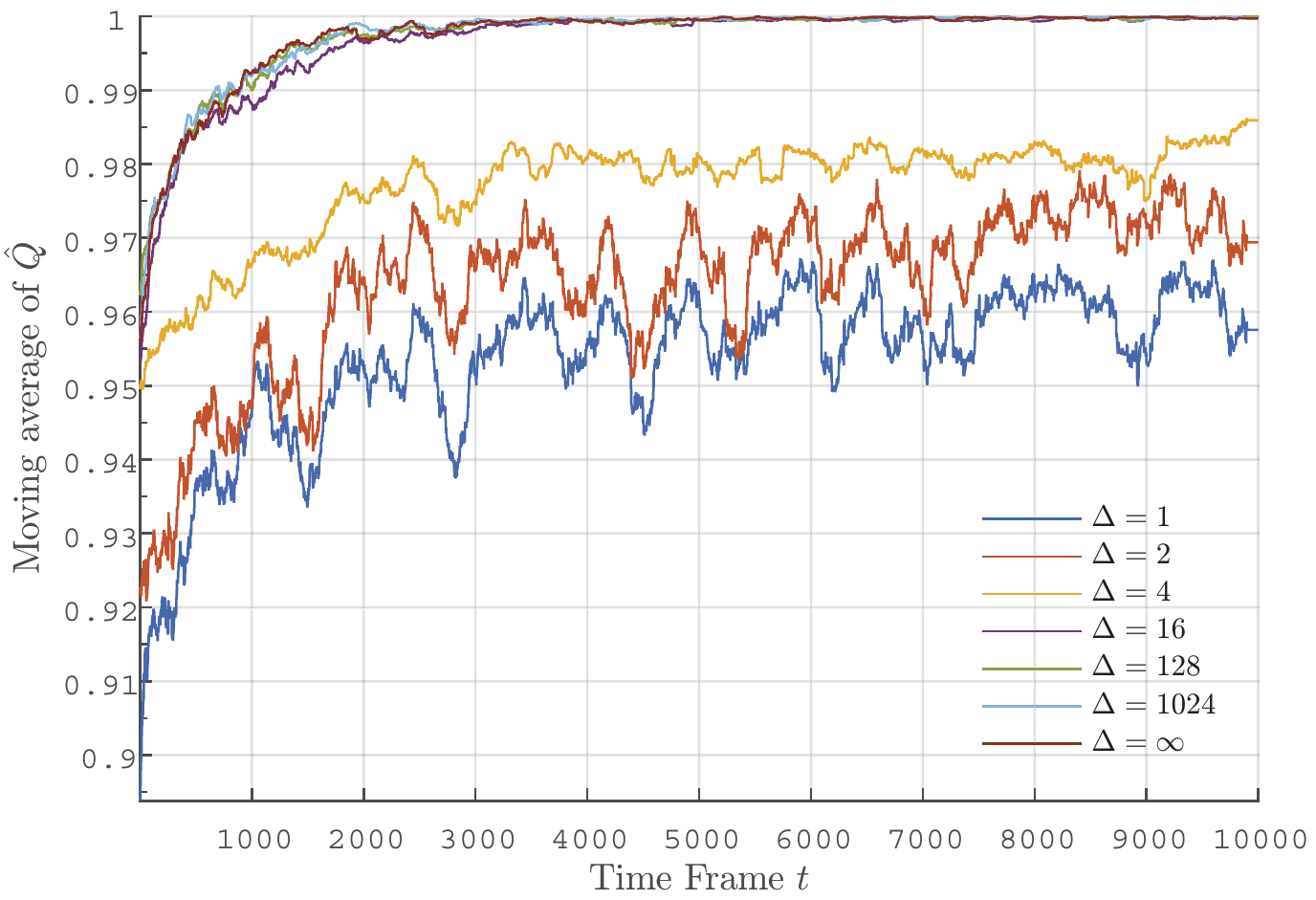}
    \end{center}
    \caption{Moving average of $\hat{Q}$ for DROO algorithm with different updating interval $\Delta$ for setting an adaptive $K$. Here, we set $N=10$.}
    \label{fig:gain_delta}
\end{figure}

\begin{figure}
\centering
  \begin{center}
    \includegraphics[width=0.47\textwidth]{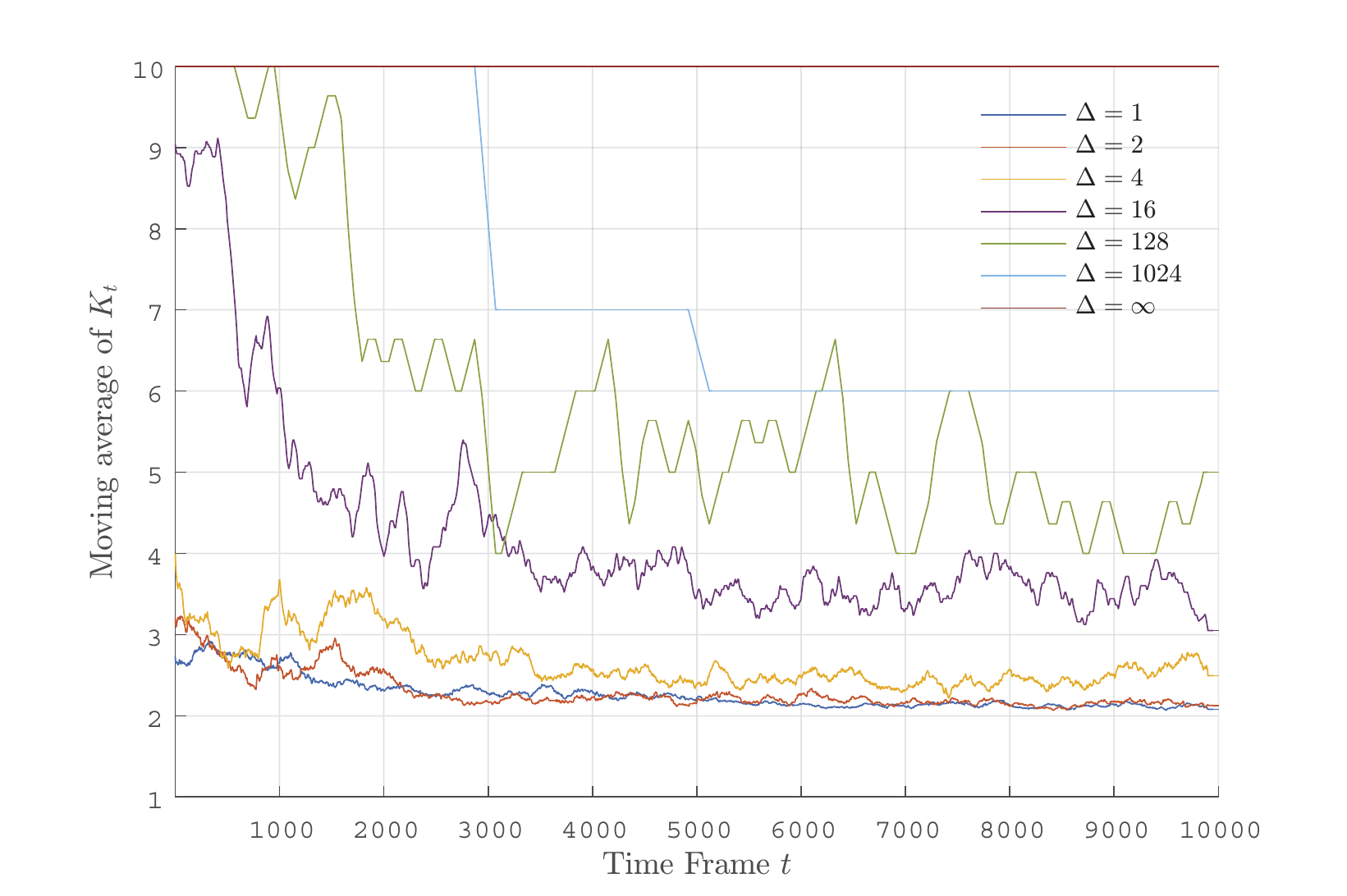}
  \end{center}
  \caption{Dynamics of $K_t$ under different updating interval $\Delta$ when $N=10$.}
  \label{fig:k_delta}
\end{figure}

In Fig.~\ref{fig:gain_delta}, we further study the impact of the updating interval of $K$ (i.e., $\Delta$) on the convergence property. Here, we use the adaptive setting method of $K$ in Section~\ref{sec:adaptive} and plot the moving average of $\hat{Q}$ over a window of $200$ time frames. We see that the DROO algorithm converges to the optimal solution only when setting a sufficiently large $\Delta$, e.g., $\Delta\geq 16$. Meanwhile, we also plot in Fig.~\ref{fig:k_delta} the moving average of $K_t$ under different $\Delta$. We see that $K_t$ increases with $\Delta$ when $t$ is large. This indicates that setting a larger $\Delta$ will lead to higher computational complexity, i.e., requires computing (P2) more times in a time frame. Therefore, a performance-complexity tradeoff exists in setting $\Delta$.

\begin{figure}
\centering
  \begin{center}
    \includegraphics[width=0.47\textwidth]{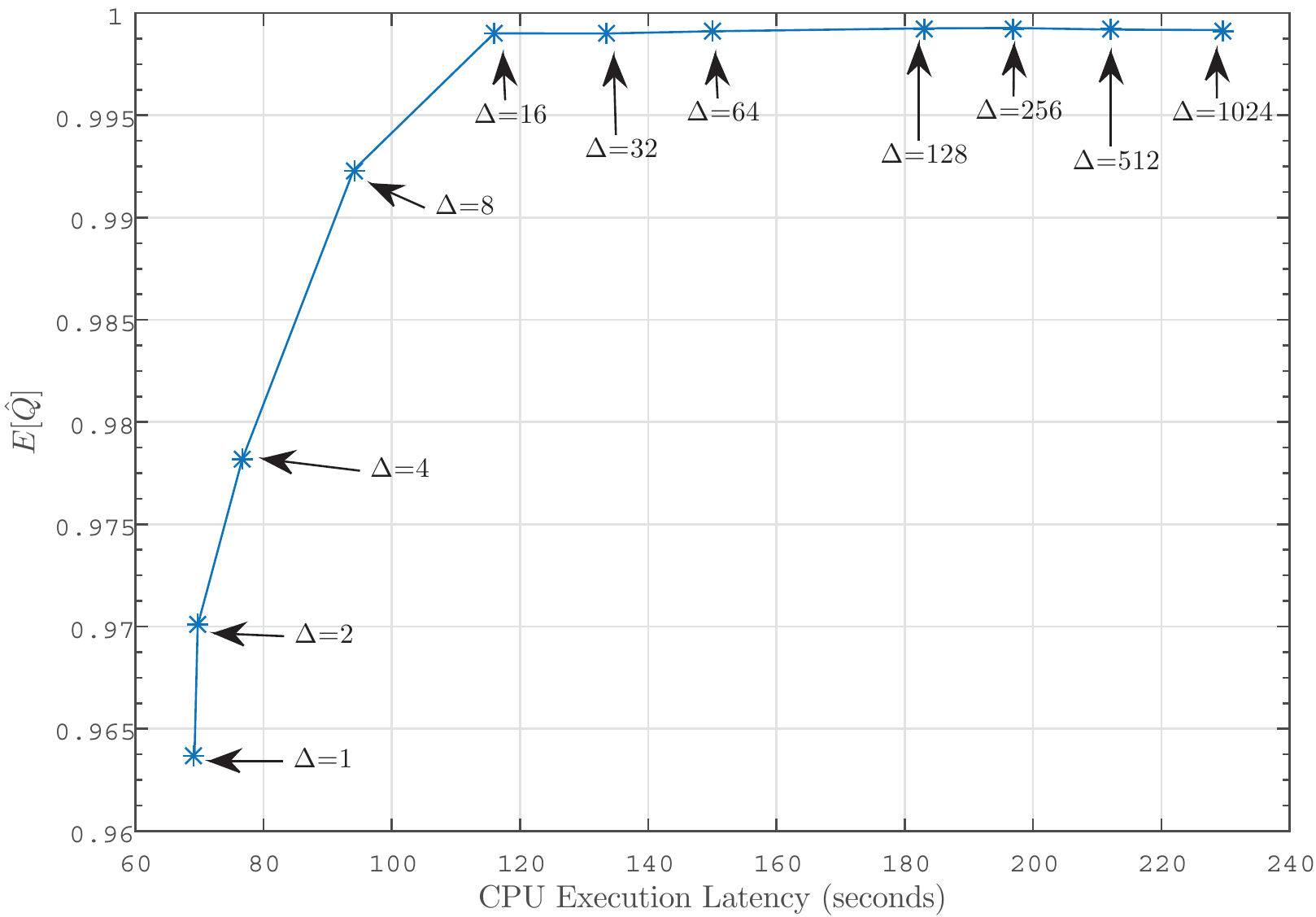}
  \end{center}
  \caption{Tradeoff between $\hat{Q}$ and CPU execution latency after training DROO for 10,000 channel realizations under different updating intervals $\Delta$ when $N=10$.}
  \label{fig:gain_time_delta}
\end{figure}

To properly choose an updating interval $\Delta$, we plot in Fig.~\ref{fig:gain_time_delta} the tradeoff between the total CPU execution latency of $10000$ channel realizations and the moving average of $\hat{Q}$ in the last time frame. On one hand, we see that the average of $\hat{Q}$ quickly increases from $0.96$ to close to $1$ when $\Delta\leq 16$, while the improvement becomes marginal afterwards when we further increase $\Delta$. On the other hand, the CPU execution latency increases monotonically with $\Delta$. To balance between performance and complexity, we set $\Delta = 32$ for DROO algorithm in the following simulations.

\subsection{Computation Rate Performance}
Regarding to the weighted sum computation rate performance, we compare our DROO algorithm with three representative benchmarks:
\begin{itemize}
	\item \textit{Coordinate Descent (CD) algorithm} \cite{bi2018twc}. The CD algorithm iteratively swaps in each round the computing mode of the WD that leads to the largest computation rate improvement. That is, from $x_i=0$ to $x_i=1$, or vice versa. The iteration stops when the computation performance cannot be further improved by the computing mode swapping. The CD method is shown to achieve near-optimal performance under different $N$.
      \item \textit{Linear Relaxation (LR) algorithm} \cite{yang2016infocom}. The binary offloading decision variable $x_i$ conditioned on (\ref{e:binary_x-i}) is relaxed to a real number between 0 and 1, as $\hat{x}_i \in [0,1]$. Then the optimization problem (P1) with this relaxed constraint is convex with respect to $\{\hat{x}_i\}$ and can be solved using the CVXPY convex optimization toolbox.\footnote{CVXPY package is online available at https://www.cvxpy.org/} Once $\hat{x}_i$ is obtained, the binary offloading decision $x_i$ is determined as follows
    \begin{align}
    x_i = \begin{cases} 1, & \textrm{when} \ r_{O,i}^*(a,\tau_i) \geq r_{L,i}^*(a), \\
    0, & \textrm{otherwise}.
    \end{cases}
    \end{align}

  \item \textit{Local Computing}. All $N$ WDs only perform local computation, i.e., setting $x_i = 0,\ i = 1,\cdots,N$ in (P2).
  \item \textit{Edge Computing}. All $N$ WDs offload their tasks to the AP, i.e., setting $x_i = 1,\ i =1,\cdots,N$ in (P2).
\end{itemize}

\begin{figure}
\centering
  \begin{center}
   \includegraphics[width=0.45\textwidth]{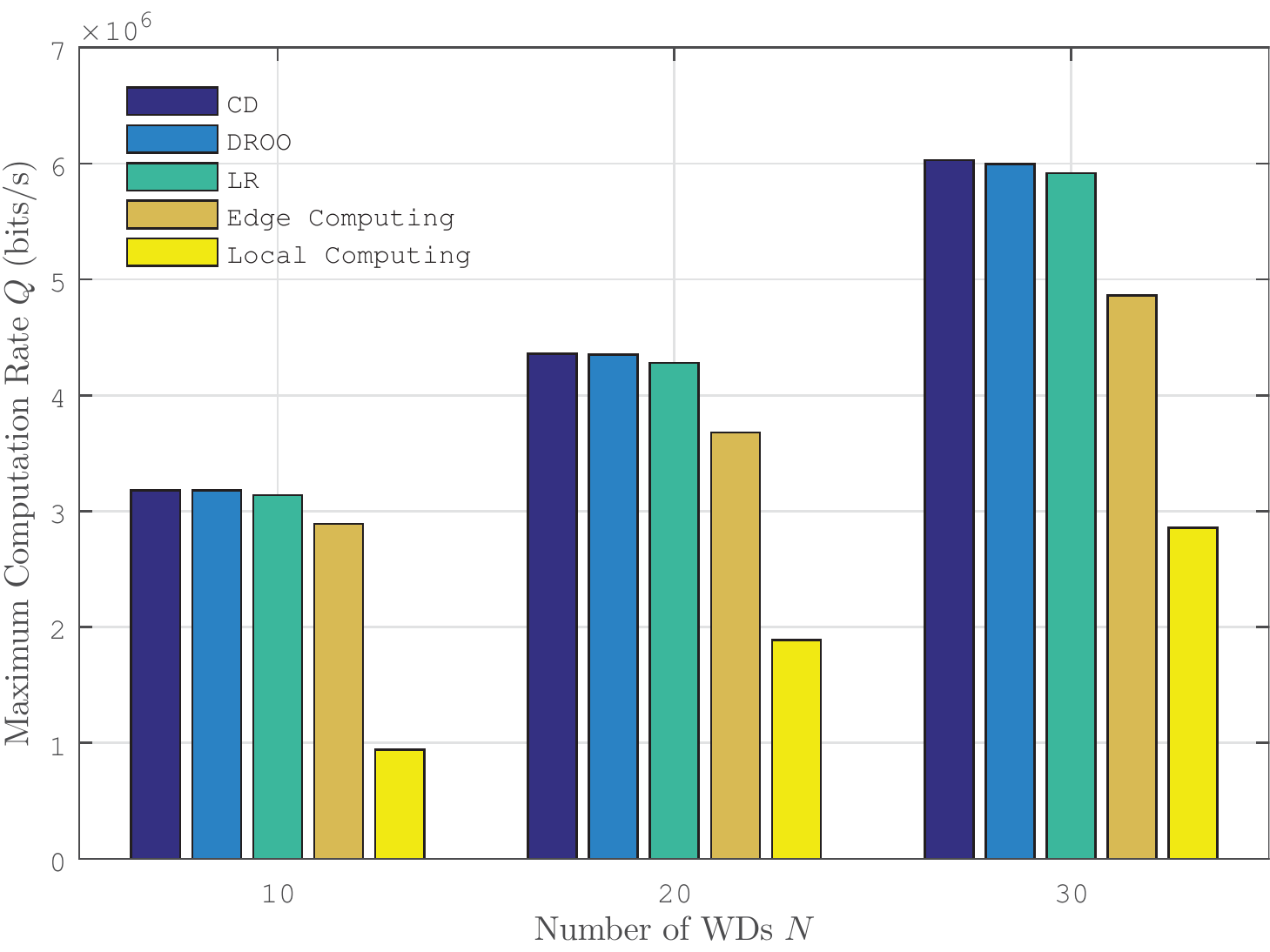}
  \end{center}
  \caption{Comparisons of computation rate performance for different offloading algorithms.}
  \label{103}
\end{figure}

\begin{figure}
\centering
  \begin{center}
   \includegraphics[width=0.45\textwidth]{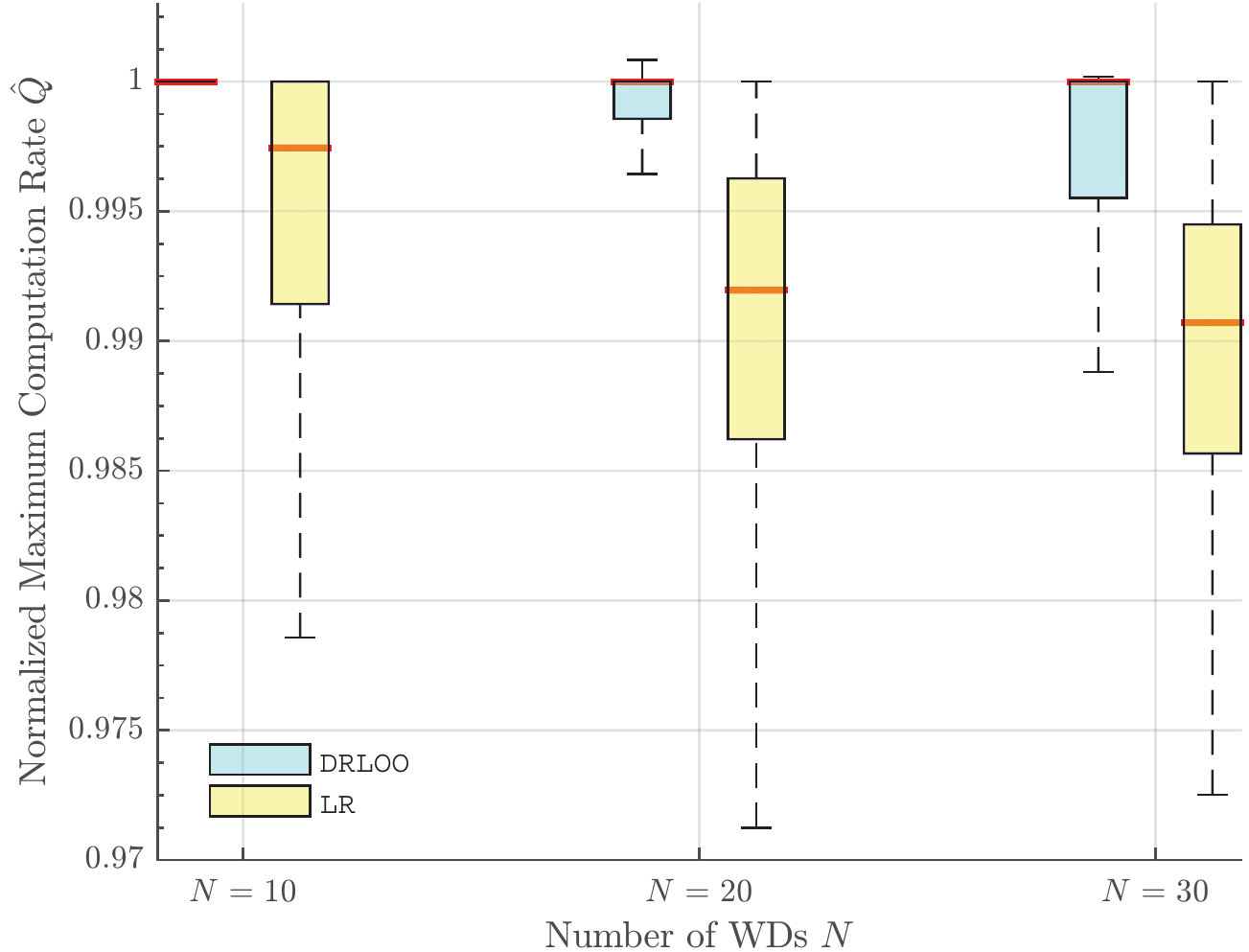}
  \end{center}
  \caption{Boxplot of the normalized computation rate $\hat{Q}$ for DROO and LR algorithms under different number of WDs. The central mark (in red) indicates the median, and the bottom and top edges of the box indicate the $25$th and $75$th percentiles, respectively. }
  \label{fig:rate_boxplot}
\end{figure}

In Fig.~\ref{103}, we first compare the computation rate performance achieved by different offloading algorithms under varying number of WDs, $N$. Before the evaluation, DROO has been trained with $24,000$ independent wireless channel realizations, and its offloading policy has converged. {This is reasonable since we are more interested in the long-term operation performance \cite{sutton2018RLbook} for field deployment.} Each point in the figure is the average performance of $6,000$ independent wireless channel realizations. We see that DROO achieves similar near-optimal performance with the CD method, and significantly outperforms the Edge Computing and Local Computing algorithms. In Fig.~\ref{fig:rate_boxplot}, we further compare the performance of DROO and LR algorithms. For better exposition, we plot the normalized computation rate $\hat{Q}$ achievable by DROO and LR. Specifically, we enumerate all $2^N$ possible offloading actions as in (\ref{e:Q_N}) when $N=10$. For $N=20$ and $30$, it is computationally prohibitive to enumerate all the possible actions. In this case, $\hat{Q}$ is obtained by normalizing the computation rate achievable by DROO (or LR) against that of CD method. We then plot both the median and the confidence intervals of $\hat{Q}$ over $6000$ independent channel realizations. We see that the median of DROO is always close-to-1 for different number of users, and the confidence intervals are mostly above $0.99$. Some normalized computation rate $\hat{Q}$ of DROO is greater than 1, since DROO generates greater computation rate than CD at some time frame. In comparison, the median of the LR algorithm is always less than 1. The results in Fig.~\ref{103} and Fig.~\ref{fig:rate_boxplot} show that the proposed DROO method can achieve near-optimal computation rate performance under different network placements.

\subsection{{Execution Latency}}
At last, we evaluate the execution latency of the DROO algorithm. The computational complexity of DROO algorithm greatly depends on the complexity in solving the resource allocation sub-problem (P2). For fair comparison, we use the same bi-section search method as the CD algorithm in \cite{bi2018twc}. The CD method is reported to achieve an $O(N^3)$ complexity. For the DROO algorithm, we consider both using a fixed $K=N$ and an adaptive $K$ as in Section~\ref{sec:adaptive}. {Note that the execution latency for DROO listed in Table~\ref{tab:computation_time} is averaged over 30,000 independent wireless channel realizations including both offloading action generation and DNN training. Overall, the training of DNN contributes only a small proportion of CPU execution latency, which is much smaller than that of the bi-section search algorithm for resource allocation. Taking DROO with $K=10$ as an example, it uses 0.034 second to generate an offloading action and uses 0.002 second to train the DNN in each time frame. Here training DNN is efficient. During each offloading policy update, only a small batch of training data samples, $|\mathcal{T}| = 128$, are used to train a two-hidden-layer DNN with only 200 hidden neurons in total via back-propagation.} We see from Table~\ref{tab:computation_time} that an adaptive $K$ can effectively reduce the CPU execution latency than a fixed $K=N$. Besides, DROO with an adaptive $K$ requires much shorter CPU execution latency than the CD algorithm and the LR algorithm. In particular, it generates an offloading action in less than $0.1$ second when $N=30$, while CD and LR take $65$ times and $14$ times longer CPU execution latency, respectively. Overall, DROO achieves similar rate performance as the near-optimal CD algorithm but requires substantially less CPU execution latency than the heuristic LR algorithm.

{The wireless-powered MEC network considered in this paper may correspond to a static IoT network with both the transmitter and receivers are fixed in locations. Measurement experiments \cite{bultitude1987measurement,howard1990Doppler,herbert2014Characterizing} show that the channel coherence time, during which we deem the channel invariant, ranges from 1 to 10 seconds, and is typically no less than 2 seconds. The time frame duration is set smaller than the coherence time. Without loss of generality, let us assume that the time frame is 2 seconds. Taking the MEC network with $N=30$ as an example, the total execution latency of DROO is 0.059 second, accounting for 3\% of the time frame, which is an acceptable overhead for field deployment. {In fact, DROO can be further improved by only generating offloading actions at the beginning of the time frame and then training DNN during the remaining time frame in parallel with energy transfer, task offloading and computation.} In comparison, the execution of LR algorithm consumes 40\% of the time frame, and the CD algorithm even requires longer execution time than the time frame, which are evidently unacceptable in practical implementation. Therefore, DROO makes real-time offloading and resource allocation truly viable for wireless powered MEC networks in fading environment.}

\begin{table*}
\caption{Comparisons of CPU execution latency}
\begin{center}
\begin{tabular}{|c| c| c| c | c | c | }
  \multirow{2}{*}{\# of WDs } &{DROO} & DROO  &   \multirow{2}{*}{CD}       &   \multirow{2}{*}{LR}    \\
                                                 & (Fixed $K = N$) & (Adaptive $K$ with $\Delta$ = 32) &  &  \\
  $10$          &   3.6e-2s &   1.2e-2s    & 2.0e-1s         &2.4e-1s      \\
  $20$          &  1.3e-1s   &  3.0e-2s   &1.3s             &5.3e-1s    \\
  $30$          &  3.1e-1s    &  5.9e-2s    &3.8s          &8.1e-1s    \\
\end{tabular}
\end{center}
\label{tab:computation_time}
\end{table*}

\section{Conclusion}\label{sec:conclusion}
In this paper, we have proposed a deep reinforcement learning-based online offloading algorithm, DROO, to maximize the weighted sum computation rate in wireless powered MEC networks with binary computation offloading. The algorithm learns from the past offloading experiences to improve its offloading action generated by a DNN via reinforcement learning. An order-preserving quantization and an adaptive parameter setting method are devised to achieve fast algorithm convergence. Compared to the conventional optimization methods, the proposed DROO algorithm completely removes the need of solving hard mixed integer programming problems. Simulation results show that DROO achieves similar near-optimal performance as existing benchmark methods but reduces the CPU execution latency by more than an order of magnitude, making real-time system optimization truly viable for wireless powered MEC networks in fading environment.

{Despite that the resource allocation subproblem is solved under a specific wireless powered network setup, the proposed DROO framework is applicable for computation offloading in general MEC networks. A major challenge, however, is that the mobility of the WDs would cause DROO harder to converge.}

{As a concluding remark, we expect that the proposed framework can also be extended to solve MIP problems for various applications in wireless communications and networks that involve in coupled integer decision and continuous resource allocation problems, e.g., mode selection in D2D communications, user-to-base-station association in cellular systems, routing in wireless sensor networks, and caching placement in wireless networks. The proposed DROO framework is applicable as long as the resource allocation subproblems can be efficiently solved to evaluate the quality of the given integer decision variables.}

\end{document}